%% file: B-L-contur.tex
\documentclass[a4paper,11pt]{article}
\pdfoutput=1
\usepackage{jheppub} 

\usepackage[T1]{fontenc} 
\usepackage{graphicx}
\usepackage{amsmath}
\usepackage{xspace}
\usepackage{subfig}
\usepackage{mydefs}
\usepackage{color}
\usepackage{float}

\usepackage{commath}
\usepackage[utf8x]{inputenc}

\title{\boldmath LHC Constraints on a $B-L$ Gauge Model using \textsc{Contur}}

\author{S. Amrith,}
\author{J. M. Butterworth,}
\author{F. F. Deppisch,}
\author{W. Liu,}
\author{A. Varma,}
\author{and D. Yallup}
\affiliation{Department of Physics and Astronomy, UCL,\\Gower Street, London WC1E 6BT, UK}

\emailAdd{shyam.amrith.14@ucl.ac.uk}
\emailAdd{j.butterworth@ucl.ac.uk}
\emailAdd{f.deppisch@ucl.ac.uk}
\emailAdd{wei.liu.16@ucl.ac.uk}
\emailAdd{abhinav.varma.17@ucl.ac.uk}
\emailAdd{david.yallup.15@ucl.ac.uk}

\abstract{The large and growing library of measurements from the Large Hadron Collider has significant power to
constrain extensions of the Standard Model. We consider such constraints on a well-motivated model involving 
a gauged and spontaneously-broken $B-L$ symmetry, within the \textsc{Contur} framework. The model contains an extra Higgs boson, 
a gauge boson, and right-handed neutrinos
with Majorana masses. This new particle content implies a varied phenomenology highly dependent on the parameters of the
model, very well-suited to a general study of this kind. We find that existing LHC measurements significantly constrain the
model in interesting regions of parameter space. Other regions remain open, some of which are within reach of future LHC data.}

\begin{document} 
\maketitle
\flushbottom

\section{Introduction}
\label{sec:intro}

The search for physics beyond the Standard Model (BSM) is one of the key motivations for the ongoing efforts at the Large Hadron Collider (LHC). Typically, signatures in a given BSM scenario, either a fully constructed ultraviolet (UV) complete model or a simplified model, are being searched through specific selection criteria optimized for a given scenario and possibly its parameters. While this approach will certainly result in the highest sensitivity, its specificity limits its application outside the initial scope. Given the large number of signatures and scenarios, it makes exploring the extensive BSM landscape a daunting task.  

In this paper we instead employ a different approach that utilizes precision measurements of Standard Model (SM) signatures. The approach, called `Constraints On New Theories Using Rivet' (\textsc{Contur}), uses particle-level differential measurements in fiducial regions of phase-space that are largely model-independent. This allows them to be compared to theoretically predicted BSM signatures. This approach, complementary to that of direct searches, can efficiently rule out BSM scenarios by comparing a large number of signatures with their measurements. The \textsc{Contur} method was introduced in \cite{Butterworth:2016sqg} where it was applied to a simplified model of Dark Matter, 
and has also been applied to exotic light scalars and to a two-Higgs-doublet model~\cite{Brooijmans:2018xbu}.
We here extend the method and apply it to a UV complete model, especially exploiting its power to test multiple signatures. 

There is significant interest in extensions to the SM in which the global symmetry behind the conservation of $B-L$ (Baryon number minus lepton number) is gauged, giving an additional $U(1)_{B-L}$ symmetry and an associated new gauge boson. 
In the specific model in which we are interested here, this additional $U(1)_{B-L}$ gauge symmetry is spontaneously broken by an extra 
SM singlet Higgs. To make the model anomaly-free it also incorporates three generations of neutral leptons sterile under the SM gauge interactions, thereby enabling the Seesaw mechanism of light neutrino mass generation.  A variant of the model was discussed in \cite{Deppisch:2018eth}, 
with a focus on signatures involving displaced vertices from relatively long-lived RH neutrinos. The parameter space of $U(1)_{B-L}$ has also been studied in previos work \cite{Das:2015nwk,Das:2016zue,Coriano:2014mpa,Coriano:2015sea,Accomando:2016sge,Accomando:2016rpc,Accomando:2017qcs,Klasen:2016qux,Ilten:2018crw,Lindner:2018kjo} with focuses on different sectors of the model. In this paper we consider the potential signatures from a wider range of model parameters and processes, in which long-lived particle decays play no significant role. We use the Herwig event generator~\cite{Bellm:2015jjp} to generate inclusively all signatures involving the new particle content of the model, and confront these expectations with LHC data using the \textsc{Contur} package~\cite{Butterworth:2016sqg} and the Rivet library~\cite{Buckley:2010ar,Bierlich:2019rhm}. This allows us to delineate regions in which LHC data already disfavour the model, and regions in which future measurements may provide sensitivity.

This paper is organized as follows. In Section~\ref{sec:blreview}, we briefly review the $U(1)_{B-L}$ gauge model and its immediate phenomenological consequences. Section~\ref{sec:theorybenchmark} then summarizes indirect theoretical considerations that constrain the relevant model parameters. In addition we here also introduce the benchmark parameter cases we study in our analysis. The important direct experimental constraints on the model are presented in Section~\ref{sec:expcons}. Our analysis using the \textsc{Contur} approach is contained in Section~\ref{sec:contur}, focussing in turn on constraints on the exotic gauge and Higgs sector. We conclude with a summary of our findings and an outlook in Section~\ref{sec:conclusions}.

\section{$B-L$ Gauge Model}
\label{sec:blreview}
In addition to the particle content of the SM, the $U(1)_{B-L}$ model contains an Abelian gauge field $B^\prime_\mu$, a SM singlet scalar field $\chi$ and three RH neutrinos $N_i$. The gauge group is $SU(3)_c\times SU(2)_L \times U(1)_Y \times U(1)_{B-L}$, where the scalar and RH neutrinos have $B-L$ charges $Y_{B-L} = +2$ and $-1$, respectively. Among the SM fields, all quarks and leptons have charges $Y_{B-L} = +1/3$ and $-1$, respectively, whereas all other SM fields are uncharged under $U(1)_{B-L}$. The scalar sector of the Lagrangian reads
\begin{align}
\label{Ls}
	{\cal L} \supset (D^{\mu}H)^\dagger(D_{\mu}H) 
	               + (D^{\mu}\chi)^\dagger D_{\mu}\chi 
	               - {\cal V}(H,\chi),
\end{align} 
with the SM Higgs doublet $H$ and the scalar potential $V(H,\chi)$ given by 
\begin{align}
\label{VHX}
	{\cal V}(H,\chi) = m^2 H^\dagger H + \mu^2 |\chi|^2 + \lambda_1 (H^\dagger H)^2 
	          + \lambda_2 |\chi|^4 + \lambda_3 H^\dagger H |\chi|^2.
\end{align}
Here, $D_\mu$ is the covariant derivative \cite{Pruna:2011me}
\begin{align}
\label{DM}
	D_\mu = \partial_{\mu} + ig_{s}\mathcal{T}_\alpha G_\mu^\alpha 
	      + igT_a W_\mu^a + ig_1 Y B_\mu + i (\tilde{g}Y + \gp Y_{B-L}) B^\prime_\mu,   
\end{align} 
where $G^\alpha_\mu$, $W^a_\mu$, $B_\mu$ are the usual SM gauge fields along with their couplings $g_s$, $g$, $g_1$ and generators $\mathcal{T}_\alpha$, $T_a$, $Y$. The Abelian gauge field $B^\prime_\mu$ couples via the $U(1)_{B-L}$ symmetry with gauge strength $g_1^\prime$ to all particles carrying a $B-L$ charge $Y_{B-L}$. In our analysis, we omit the Abelian mixing between $U(1)_{B-L}$ and $U(1)_{Y}$, $\tilde{g} = 0$, as a simplification. This means we consider the minimal gauged $B-L$ model. Consequently, the SM gauge sector is extended to include the kinetic term 
\begin{align}
\label{LYM}
	{\cal L} \supset -\frac{1}{4} F^{\prime\mu\nu} F_{\mu\nu}^\prime,
\end{align} 
with the field strength tensor of the $B-L$ field, $F^\prime_{\mu\nu} = \partial_\mu B^\prime_\nu - \partial_\nu B^\prime_\mu$. This is manifest observationally as a new gauge boson, \ZP, coupling to SM fermions with a characteristic coupling \gp.

The fermion part of the Lagrangian now contains a term for the right-handed neutrinos
\begin{align}
\label{Lf}
	{\cal L} \supset i\overline{\nu_{Ri}}\gamma_\mu D^\mu \nu_{Ri}, 
\end{align} 
but is otherwise identical to the SM apart from the covariant derivatives incorporating the $B-L$ gauge field. Here, a summation over the fermion generations $i = 1, 2, 3$ is implied. Finally, the Lagrangian contains the additional Yukawa terms
\begin{align}
\label{LY}
	{\cal L} \supset 
	   - y_{ij}^\nu \overline{L_i}\nu_{Rj}\tilde{H}
	   - y_{ij}^M   \overline{\nu^c_{Ri}} \nu_{Rj}\chi 
	   + \text{h.c.},
\end{align} 
where $L$ is the SM lepton doublet, $\tilde{H} = i\sigma^2 H^\ast$ and a summation over the generation indices $i, j = 1, 2, 3$ is implied. The Yukawa matrices $y^\nu$ and $y^M$ are at this point general $3\times 3$ matrices; RH neutrino masses \MNR are generated by the breaking of the $B-L$ symmetry through the vacuum expectation value $x = \langle\chi\rangle$ as outlined below, with the mass matrix given by $M_R = \sqrt{2}y^M x$. The light neutrinos mix with the RH neutrinos via the Dirac mass matrix $m_D = y^\nu v/\sqrt{2}$, generated after EW symmetry breaking, $v = \langle H^0 \rangle$. The complete mass matrix in the $(\nu_L, \nu_R^c)$ basis is then
\begin{align}
\label{MD}
	{\cal M} = 
	\begin{pmatrix}
		0   & m_D \\
		m_D & M_R
	\end{pmatrix},
\end{align} 
In the well studied seesaw limit, $M_R$ $\gg$ $m_D$, the light and heavy neutrino masses are $m_\nu \sim - m_D M^{-1}_{R} m^T_D$ and $M_N \sim M_R$. The flavour and mass eigenstates of the light and heavy neutrinos are connected as  
\begin{align}
\label{Neutrino}
	\begin{pmatrix}
		\nu_L \\ \nu_R
	\end{pmatrix} = 
	\begin{pmatrix}
		V_{LL} & V_{LR} \\
		V_{RL} & V_{RR}
	\end{pmatrix}
	\begin{pmatrix}
		\nu \\ N
	\end{pmatrix},
\end{align} 
schematically written in terms of 3-dimensional blocks in generation space. The SM charged current lepton mixing $V_{LL} = U_\text{PMNS}$ is determined by oscillation experiments (in the basis of diagonal charged lepton masses and apart from small non-unitarity corrections ) whereas the active-sterile mixing $V_{LR} \lesssim 0.1 - 0.01$ is constrained by electroweak precision data, largely independent of the RH neutrino mass. More stringent but highly mass-dependent constraints can be set from direct searches at the LHC, lepton colliders and high intensity experiments, see \cite{Deppisch:2015qwa} and references therein. For the simplifying case a single generation of light and heavy neutrinos we will consider, Eq.~\eqref{Neutrino} reduces to the $2\times 2$ form 
\begin{align}
\label{Rotation}
	\begin{pmatrix}
		\nu_L \\ \nu_R
	\end{pmatrix} = 
	\begin{pmatrix}
    	\cos\theta_\nu & -\sin\theta_\nu \\
		\sin\theta_\nu &  \cos\theta_\nu
	\end{pmatrix}
	\begin{pmatrix}
		\nu \\ N
	\end{pmatrix}.
\end{align}
For simplicity, we thus neglect mixing among flavours and therefore generations decouple. This corresponds to a diagonal Yukawa coupling matrix $y^\nu_{ii} = \sqrt{2} M_{N_i} V_{iN}/v$ with $i = e, \mu, \tau$ and using the neutrino seesaw relation. Here, $V_{iN}$ represents the active-sterile mixing, $\sin\theta_i = V_{iN}$, in the three generations. 

Crucial for the above to work, the $U(1)_{B-L}$ symmetry is broken by the vacuum expectation value of the additional scalar singlet $\chi$ which then also causes it to mix with the SM Higgs. The mass matrix of the Higgs fields $(H,\chi)$ at tree level is \cite{Robens:2015gla}
\begin{align}
\label{mass}
	M_h^2 = \begin{pmatrix}
		2\lambda_1 v^2 & \lambda_3 x v \\
		\lambda_3 x v & 2\lambda_2 x^2
	\end{pmatrix}.
\end{align}
The physical masses of the two Higgs states $h_1, h_2$ are then 
\begin{align}
\label{Higgsmass}
	M^2_{h_{1(2)}} = \lambda_1 v^2 + \lambda_2 x^2 
	-(+) \sqrt{(\lambda_1 v^2 - \lambda_2 x^2)^2 + (\lambda_3 xv)^2},
\end{align} 
and the ($h_1,h_2)$ states are related to the gauge states ($H, \chi$) via
\begin{align}
\label{Higgs mixing}
	\begin{pmatrix}
		h_1 \\ h_2
	\end{pmatrix} = 
	\begin{pmatrix}
    	\cos\alpha & -\sin\alpha \\
		\sin\alpha &  \cos\alpha
	\end{pmatrix}
	\begin{pmatrix}
		H \\ \chi
	\end{pmatrix}.
\end{align} 
The directly measurable parameters for the Higgs sector are the masses $M_{h_1}$ and $M_{h_2}$, as well as the mixing angle $\alpha$ expressed as
\begin{align}
\label{lambda}
	\tan(2\alpha) = \frac{\lambda_3 v x}{\lambda_2 x^2 - \lambda_1 v^2}.
\end{align}

The other measurable independent parameters can be taken to be the mass and coupling of the \ZP, \MZP, \gp, and the RH neutrino masses \MNR.

The $U(1)_{B-L}$ model is phenomenologically appealing. With the inclusion of three copies of right-handed neutrinos it is an anomaly-free gauge theory that incorporates three different simplified scenarios through mixing with the SM singlets of the model: the $Z'$ via its mixing with the SM $Z$, the extra Higgs $h_2$ mixing with the SM Higgs and the right-handed neutrinos $N_1$ mixing with the active neutrinos. Formally, any two of the above can be switched off by taking an appropriate limit to yield a simplified model with only an extra \ZP, a singlet scalar or a singlet neutrino. These scenarios have been studied extensively in the literature. For example, an easy way to achieve a simplified model with only singlet neutrinos is to assume a very high $U(1)_{B-L}$ breaking scale (with vanishing Higgs and gauge mixing), but make the heavy neutrino Yukawa couplings $y^M$ small to keep the states accessible. 

\section{Theoretical Constraints and Benchmark Scenarios}
\label{sec:theorybenchmark}

While we focus on direct experimental constraints from LHC searches in this work, we still need to incorporate theoretical considerations and indirect experimental constraints to disregard parameter space that is either unphysical or where a perturbative treatment is not possible. Theoretical considerations may also hint at interesting parameter space. Below, we discuss the requirement of vacuum stability and perturbativity, both at the EW scale as well as higher scales. As the model incorporates new exotic particles, it can also affect electroweak precision observables. The most sensitive quantity in this regard is the SM $W$ boson mass, which we 
consider below. We here omit constraints from perturbative unitarity which will set upper limits on the extra Higgs mass but which are generally less severe than the constraints considered above \cite{Pruna:2011me}.

\subsection{Vacuum Stability and Perturbativity}
\label{sec:vac}

A basic requirement is that the vacuum is stable; this puts a constraint on the parameters in the scalar potential. As we would like to impose these constraints on observable quantities, we express the couplings for the quartic terms of the Higgs potential in Eq.~\eqref{Higgs mixing} as 
\begin{align}
\label{para_quartic}
	\lambda_1 &= 
 		\frac{1}{4 v^2} [(M_{h_1}^2 +\MHH^2) - 
    	\cos 2 \alpha (\MHH^2 - M_{h_1}^2)], \nonumber\\
	\lambda_2 &= 
 		\frac{1}{4 x^2} [(M_{h_1}^2 +M_{h_2}^2) + 
    	\cos 2 \alpha (\MHH^2 - M_{h_1}^2)], \\
	\lambda_3 &= 
 		\frac{1}{2 v x} 
    	[\sin 2 \alpha (\MHH^2 - M_{h_1}^2)] \nonumber
\end{align} 
The vacuum stability condition then requires that \cite{Chakrabortty:2013zja}
\begin{align}
	4 \lambda_1 \lambda_2 - {\lambda_3}^2  > 0, \quad 
	\lambda_1 > 0, \quad
	\lambda_2  > 0.
\end{align}
In addition, perturbativity requires the couplings in the model to be small enough such that loop corrections remain bounded. We choose the upper limit conservatively to be $\abs{\lambda_{1,2,3}} < 1$.
Our chosen model parameters are \MZP, \gp, \MHH, \SA, \MNR and $V_{lN}$. Among these, $M_N$ and $V_{lN}$ do not enter to the scalar vacuum stability and perturbative constraints and we display the allowed region for each pair of parameters when setting the other pair of parameter to be a constant or some other reasonable relation. In addition to the scalar parameters, the gauge and fermion Yukawa couplings also need to remain perturbative but this simply means we restrict the relevant parameters to be $\gp < 1, \MNR/x < 1$.

\subsection{Renormalisation Group Evolution}
\label{sec:rge}

We input parameters as shown in the Table~\ref{tab:scenarios} at electroweak scale, then evolve all model parameters according to their respective renormalisation group equations (RGEs). Requiring that the model remains well-defined and perturbative at higher energy scales $Q > Q_{\text{EW}}$ puts additional constraints on the parameter space. If we were to assume that the $B-L$ is the `ultimate' theory, i.e. not superseded by a new model at some scale $Q_\text{UV}$, we should require vacuum stability and perturbativity all the way up to the Planck scale. This is of course a very strong theory bias and we only use it to highlight potentially interesting parameter space. In our calculations, we use the RGEs in the $B-L$ model given in \cite{Chakrabortty:2013zja} which is shown in the appendix \ref{sec:app} .

\subsection{$W$ Boson Mass Constraint}

An additional indirect constraint arises from the shift of the $W$ boson mass via radiative effects of the extra Higgs in the model. This is quantified by a parameter
$\Delta r$ relating Fermi constant $G_F$, the fine structure constant $\alpha_{EM}$ and the electroweak renormalised gauge boson masses $m_Z$, $m_W$ \cite{Lopez-Val:2014jva},
\begin{align}
		m_W^2 \left(1-\frac{m_W^2}{m_Z^2}\right) = 
		\frac{\pi\alpha_{EM}}{\sqrt{2}G_F}(1+\Delta r).
\end{align}
In the SM, $\Delta r = 0.038$ which gives $m_W = 80.360$~GeV, compared to the tree level value $m_W^\text{tree} = 80.94$~GeV, with a theoretical uncertainty of around 4~MeV \cite{Lopez-Val:2014jva}. However, the experimental data gives $m_W^\text{exp} = 80.385 \pm 0.015$~GeV \cite{Patrignani:2016xqp}, which is therefore somewhat in tension with the SM prediction.

Extra particles in BSM scenarios can contribute to the mass shift. In our scenario, the singlet Higgs does so with $\Delta r = \Delta r_\text{SM} + \delta(\Delta r)$ leading to a mass shift
\begin{align}
	\Delta m_W = -\frac{1}{2} m_W 
	\frac{\sin^2\theta_W}{\cos^2\theta_W - \sin^2\theta_W}\delta(\Delta r).
\end{align}
The extra contribution $\delta(\Delta r)$ involves Higgs loops which are dependent on the masses of the two Higgs particles and their mixing angle $\alpha$. The above discrepancy between the SM prediction and the observed $W$ boson mass could be resolved if the extra Higgs is lighter than the SM Higgs \cite{Lopez-Val:2014jva}. We omit this possibility and instead use the above to set a constraint on the $M_{h_2}$ - $\sin\alpha$ parameter space by requiring that the calculated $m_W$ is within 2$\sigma$ of its experimental value as described in \cite{Robens:2015gla}.

\subsection{Benchmark Scenarios}
\label{sec:benchmarks}
There are six extra free parameters compared to the SM which can be categorised into three pairs: \MZP and \gp describing the gauge sector and also fixing the vacuum expectation value (vev) of the $B-L$ gauge; \MHH and \SA describing the extra Higgs mass eigenstate and the mixing between the two Higgs fields; and similarly \MNR and $V_{lN}$ for the heavy neutrino and its mixing strength with the active neutrinos. As we will describe in the next section, we largely leave aside channels incorporating heavy neutrinos, either because the discovery signal is too small as the upper bound on the neutrino mixing is $V_{lN} \lesssim 10^{-2}$ \cite{Khachatryan:2015gha}, or the heavy neutrino decays in a displaced vertex which cannot be captured by the \textsc{Contur} analysis in the following discussion. Thus, we safely set the neutrino masses to be $\MNR = \MZP / 5$ in all cases. This ensures that heavy neutrino Yukawa couplings are always smaller than $\gp$. We choose $V_{lN}$ as determined by the Type-I seesaw generation of light neutrino masses, $V_{lN} = \sqrt{\frac{m_\nu}{\MNR}}$ where $m_\nu = 0.1$~eV is the mass scale of light neutrinos.

\begin{table}[h]
	\centering	
	\begin{tabular}{c|cccccc}
		\hline
		Scenario & $\MZP$ [GeV] & $\gp$ & $M_{h_2}$ & $\sin \alpha$ & $M_{N_i}$ \\
		\hline
		A & [1, $10^4$] & [$3\times 10^{-5}$, 0.6] & $\MZP/(2\gp)$ & 0   & $\MZP/5$ \\
		B & [1, $10^4$] & [$3\times 10^{-5}$, 0.6] & $\MZP/(2\gp)$ & 0.2 & $\MZP/5$ \\
		C & [1, $10^4$] & [$3\times 10^{-5}$, 0.6] & 200 GeV       & 0.2 & $\MZP/5$ \\
		\hline
		D & 7000 & 0.2       & [0, 800]~GeV & [0, 0.7] & $\MZP/5$ \\ 
		E & 35   & $10^{-3}$ & [0, 800]~GeV & [0, 0.7] & $\MZP/5$ \\
		\hline
	\end{tabular}
	\caption{Benchmark scenarios used in our analysis. In addition, the active-sterile neutrino mixing is fixed as $V_{lN} = \sqrt{0.1~\text{eV} / \MNR}$, independent of the generation of the heavy neutrino.}
	\label{tab:scenarios}
\end{table}
We thus focus on the other parameters where we will scan over two-dimensional slices of the parameters while keeping the other two parameters fixed. Our parameter choices are summarised in Table~\ref{tab:scenarios}. For \MZP and \gp, we can choose to switch off the effects of the Higgs mixing and the second Higgs mass eigenstates by setting $\SA = 0$ and $\MHH = \MZP/(2\gp)$  the $B-L$ gauge breaking vev $x$ (Case A). We can also consider only the Higgs mixing by setting $\SA = 0.2$ (Case B) which is still allowed by the direct experimental limits described below and the $W$ mass constraint \cite{Lopez-Val:2014jva}. Finally, we can switch on both, setting $\SA = 0.2$ and $\MHH = 200$~GeV (Case C); while still allowed from theoretical considerations and Higgs property determinations, this choice will have stronger constraints from searches.

In choosing \MHH and \SA, as the $B-L$ breaking scale needs to be higher than 3.45~TeV experimentally as described below, we use $\gp = 0.2$ and $\MZP = 7$~TeV (Case D). Another parameter choice (Case E) can be proposed when we combine the effects of a light $\ZP$ by setting $\MZP = 35$~GeV with $\gp = 10^{-3}$ yielding the same vev as Case D. This would allow a possible production channel of heavy neutrinos as many more $\ZP$ are produced, due to its light mass and subsequent decays to the lighter heavy neutrinos.

\begin{figure}[t!]
	\subfloat[]{\includegraphics[width=0.51\textwidth]{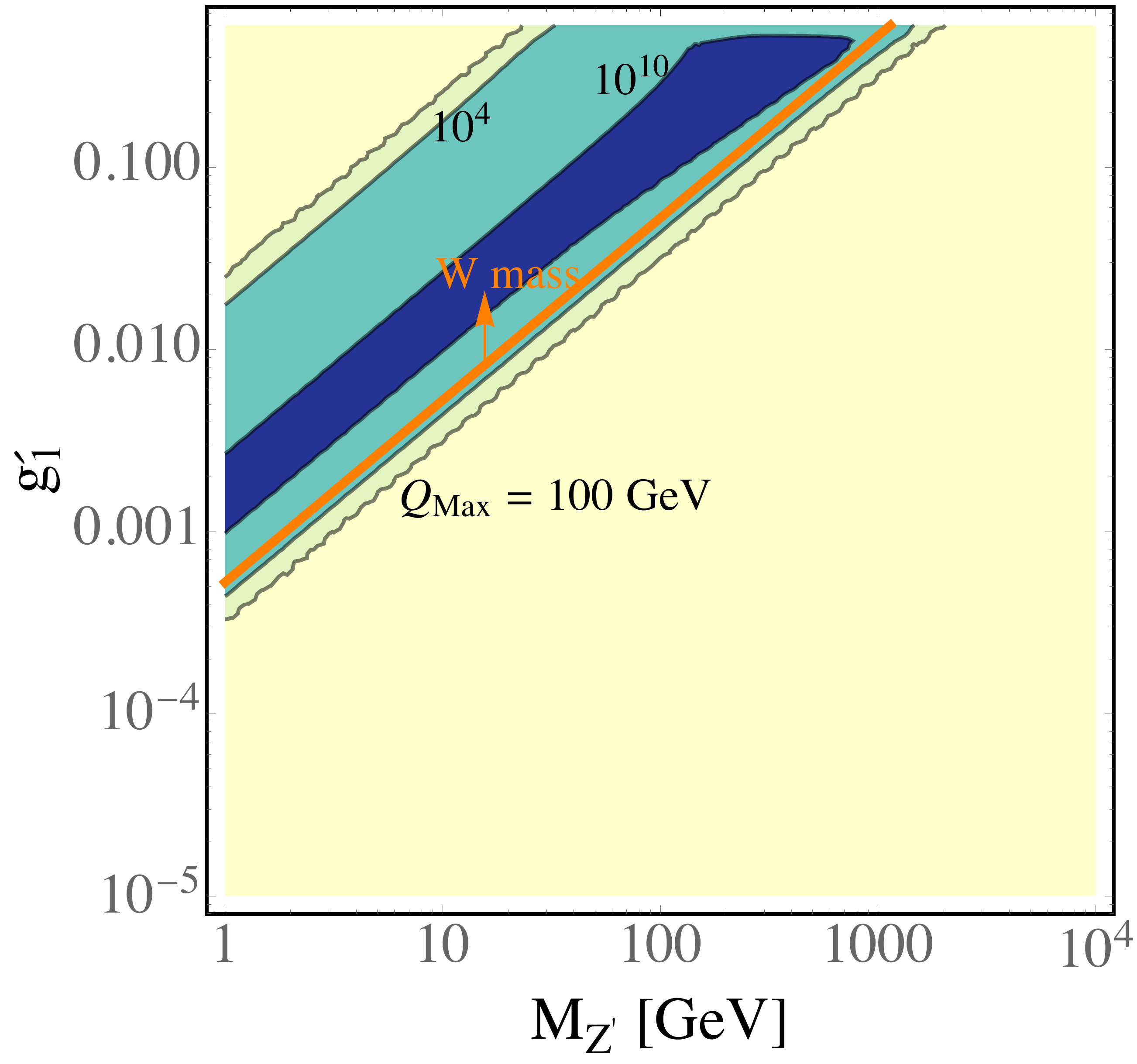}}
	\subfloat[]{\includegraphics[width=0.49\textwidth]{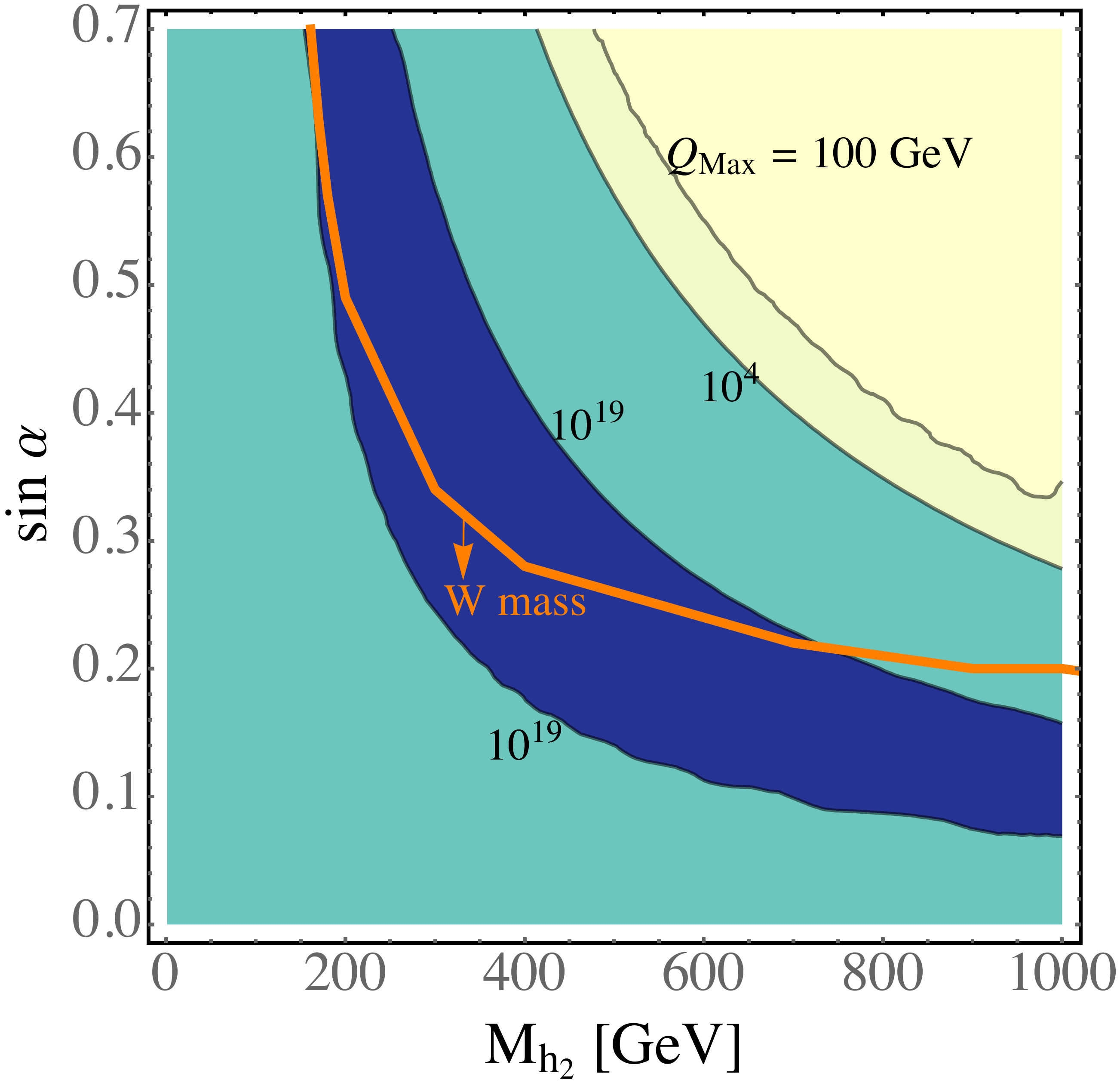}}
	\caption{Maximal perturbative scale $Q_{\text{Max}}$ in GeV and constraint from electroweak $W$ mass corrections as a function of (a) $\gp$ and $\MZP$ with $\MHH = \MZP/(2\gp)$ and \SA = 0.2 (Case B) and 
(b) $M_{h_2}$ and $\SA$ with $\MZP = 7$~TeV and $\gp = 0.2$ (Case D). The $W$ mass constraint is satisfied above (below) the depicted contour in panel a (b), as indicated by the arrows.}
	\label{fig:rgescale}
\end{figure}
In Fig.~\ref{fig:rgescale}, we show the maximal scale $Q_{\rm Max}$ up to which the model remains perturbative for Case B and D, as a function of the respective running model parameters. In Fig.~\ref{fig:rgescale}~(a), only a narrow band from $\gp\approx 10^{-3}$, $\MZP \approx 1$~GeV to $\gp\lesssim 1$, $\MZP \approx 1$~TeV permits $Q_\text{Max}$ as high as $10^{10}$ GeV. This is an indirect effect as the extra Higgs mass is adjusted as $M_{h_2} = \frac{\MZP}{2\gp}$ while $\sin\alpha$ = 0.2 is kept at a constant and fairly high value. This behaviour becomes clearer for Case D, depicted in Fig.~\ref{fig:rgescale}~(b). Here, perturbativity rules out simultaneously large \MHH and \SA but it permits a hyperbolic band where $Q_{\rm Max}$ is at or above the Planck scale. In addition, the constraint from the $W$ boson mass corrections is shown in both plots as well. In Fig.~\ref{fig:rgescale}~(a), the allowed region from this consideration is \emph{above} the depicted line while in Fig.~\ref{fig:rgescale}~(b), the region \emph{below} the corresponding line is allowed.

\section{Existing Experimental Constraints}
\label{sec:expcons}

Before analyzing the constraints from LHC SM searches using the \textsc{Contur} framework, we  here briefly summarize other experimental constraints on the model parameters - in particular, the additional gauge boson mass \MZP, the $B-L$ gauge coupling \gp, the Higgs mixing angle \SA, the second Higgs mass \MHH and the RH neutrino mass \MNR.

For \MZP and \gp, the experimentally sensitive parameter is the vev of the $B-L$ gauge-breaking Higgs. Resonance searches in $pp\to \ZP \to l^+l^-$ bound the \ZP mass to $\MZP \gtrsim 4.5$~TeV \cite{Aaboud:2017buh,Sirunyan:2018exx} with a SM-valued gauge coupling. Searches at LEP-II \cite{Cacciapaglia:2006pk, Anthony:2003ub, LEP:2003aa, Carena:2004xs} for a resonance constrain \ZP mass and gauge coupling, and thus the $B-L$ breaking scale $x \equiv \MZP/(2\gp) \geq 3.45$~TeV. Thus for Cases D and E, the vev 17.5~TeV we have selected is allowed. For Case A, this limit lies within the region subsequently excluded by LHC dilepton searches.

A constraint applicable to the whole range of $\ZP$ masses we consider here is provided through the measurement of the electron-neutrino cross section 
principally from Charm II~\cite{Lindner:2018kjo,Vilain:1993kd}, obtained via the  \emph{Darkcast} framework~\cite{Ilten:2018crw}.

Searches for dark photons can be recast to other BSM models, and the \emph{Darkcast} framework can use these to provide the corresponding limits for the $B-L$ model~\cite{Ilten:2018crw}. This extends the current experimental limits on \gp for low \MZP, such that $\gp < 10^{-4}$ for \MZP < 10 GeV and $\gp \lessapprox 10^{-3}$ for 10 GeV~$ < \MZP < 70$~GeV. The latter region is dominated by the LHCb dark photon search~\cite{Aaij:2017rft}. 
As the Higgs mixing \SA is not considered in the production mechanisms, these limits cannot be directly applied to Case B or C, although they can be expected to have some impact.

The SM singlet Higgs and its mixing angle $\alpha$ with the SM Higgs are constrained by perturbativity and unitarity considerations \cite{Pruna:2011me}, setting an upper limit on \MHH as described above. Additionally, direct searches at the LHC for a BSM Higgs signal further constrain the mixing such that 
$\SA \lesssim 0.35$ in the aforementioned mass range \cite{CMS:xwa}. An indirect constraint on the Higgs mixing angle $\sin^2 \alpha \lesssim 0.31$ can also be obtained from  the measurement of SM Higgs decays into a number of SM final states \cite{Khachatryan:2014jba, Banerjee:2015hoa}. The bound coming from SM Higgs signal strength measurement is valid for all masses of the BSM Higgs $\MHH > M_{h_1}$. 

In the present work, we consider relatively low mass right-handed neutrinos, $M_N = \MZP/5$, in order to ensure there are decay channels open to the \HH in all scenarios. This means that RH neutrinos may be pair-produced from \ZP decays. In a pure Type-I seesaw scenario, the RH neutrino mass is related to the mixing with the active neutrinos and the light neutrino mass via $\MNR \sim \frac{m_\nu}{V_{lN}^2}$. The sub-eV scale light neutrino mass constraints from $0\nu\beta\beta$ and Tritium beta decay experiments as well as from cosmological observations such as Planck \cite{Dev:2012bd, Ade:2015xua} together with the maximal active-sterile mixing $V_{lN} \sim 0.01$ limited by direct searches (see e.g. \cite{Deppisch:2015qwa} and references therein) gives a lower limit $\MNR > 1$~keV which is easily satisfied in the \MZP region we choose.

\section{LHC Constraints Using \textsc{Contur}}
\label{sec:contur}
Our analysis proceeds as follows. 
The Lagrangian for the model is coded in Feynrules~\cite{Alloul:2013bka} and used to produce a UFO~\cite{Degrande:2011ua} file,
which is then read into Herwig7~\cite{Bellm:2015jjp}\footnote{Version 7.1.4, changeset 0d744493e50e is used for the limit plots. 
Version 7.1.2 is used for the RH neutrino lifetime only.}. 
All tree-level processes involving one of more BSM particles 
($N, \ZP, h_2$) in the matrix element are generated in proton-proton collisions at 7, 8 and 13 TeV. 
The effective Higgs couplings to gluons and photons via loops are also included. Any interference terms between BSM
and SM contributions to final states are neglected.
Unstable particles are decayed by Herwig. QCD and QED radiation are simulated using a leading-logarithmic shower.
Underlying event,  hadronisation and hadron decays are also simulated, to produce a realistic event final state.
These events are passed to Rivet~\cite{Buckley:2010ar}, so that their contribution to the fiducial regions of LHC measurements, 
from ATLAS, CMS and LHCb, can be evaluated.
This is possible because the measurements are defined in terms of idealised 
final-state particles and corrected for detector effects.
This is done within fiducial regions in which the detector has high acceptance, a procedure which minimises model dependence.
The significance of the additional BSM contributions relative to the experimental uncertainties on the measurements is used to derive
a confidence level (CL) at which the given BSM parameter point is disfavoured, on the assumption that the measurement is equal to the SM. This comparison
is made using \textsc{Contur}~\cite{Butterworth:2016sqg} and is roughly equivalent to treating all the measurements (which have been shown to
agree with the SM) as data-driven control regions. A simple $\chi^2$ test statistic is used, the confidence interval is calculated for the different signal hypotheses using asymptotic distributions of the test statistic~\cite{Cowan:2010js} and interpreted as a CL using the CL$_{s}$ formalism~\cite{Read:2002hq}. Statistical correlations are eliminated by only taking the
most significant point from any data set where there are overlaps of events. Systematics are assumed to be 100\% correlated within a distribution
and uncorrelated between distributions.

\subsection{Exotic Production and Decay Modes}

\begin{figure}[h]
\subfloat[]{\includegraphics[width=0.5\textwidth]{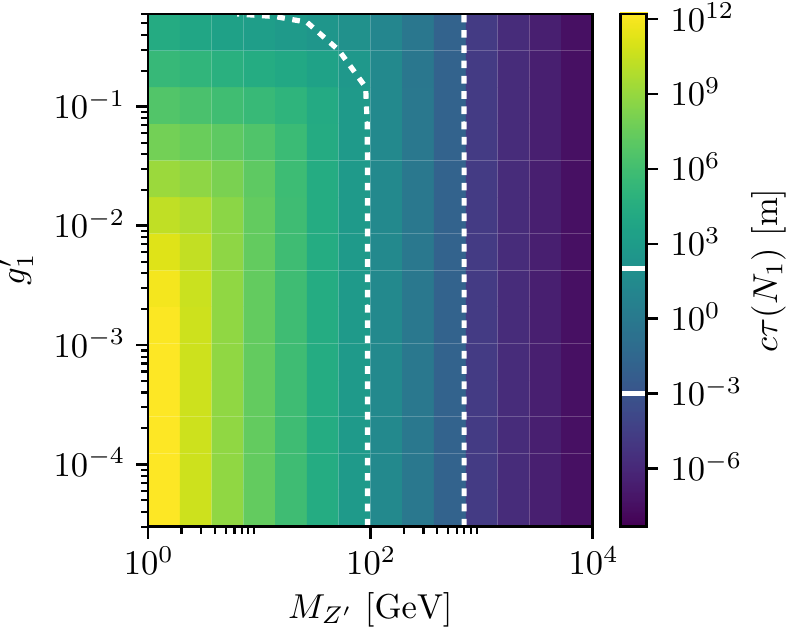}}
\subfloat[]{\includegraphics[width=0.5\textwidth]{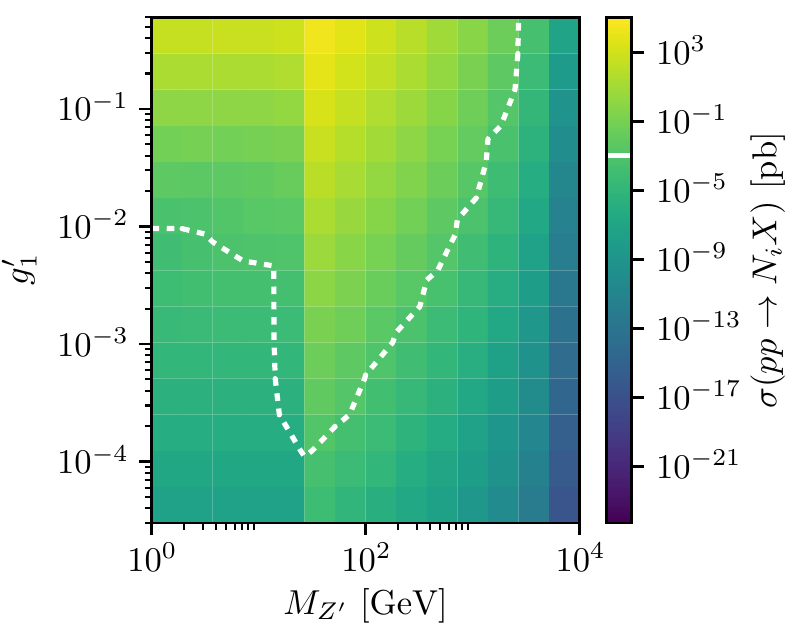}}
\caption{(a) The proper decay length of the heavy RH neutrino for Case B. The dashed lines indicate the boundaries of region between 
100~m$~> c\tau >$~1~mm within which the neutrino would manifest a ``long-lived particle'' signal.
(b) the total production cross section for the RH neutrino in Case B, for 8~TeV $pp$ collisions. The dashed line 
indicates the 1~fb contour,
corresponding to roughly 30 events before any cuts, for the maximum luminosity considered here.
}
\label{fig:neutrinos}
\end{figure}

The relevant production and decay modes vary depending upon the parameters of the model. However, the most important are 
\begin{itemize}
\item the direct production of \ZP, (or for lower masses, multiple \ZP), often in association with hadronic jets, and with subsequent \ZP decay to leptons
\item the decay of the SM Higgs to \ZP pairs.
\item production of the \HH via gluon fusion, with subsequent decay to weak bosons.
\item production of the \HH via gluon fusion, with subsequent decay to \ZP.
\item associated production of the \HH or \ZP with $\gamma$, $W$ or $Z$.
\end{itemize}
The RH neutrino masses are all set to $\MNR = \MZP/5$, and these neutrinos may also be produced. 
Their proper decay length $c\tau \approx 2.5 \times 10^{5}\ \text{m} \times (\frac{1~{\rm GeV}}{\MNR})^4$, for $1~{\rm GeV} \ll \MNR < M_W$,
will vary across the parameter space, as shown as an example for Case B 
in Fig.~\ref{fig:neutrinos}a. 
For low \MZP (and hence low \MNR), $c\tau > 10$m and the neutrinos may be considered to be stable for our purposes.
However as \MZP increases above the $Z$ mass, $c\tau$ decreases, leading first to displaced decays within the detector volume, 
and eventually to effectively prompt decays.
The \textsc{Contur} approach is not well-suited to considering displaced vertex decays, 
since the fiducial cross section definitions and the detector 
corrections applied to obtain them typically consider only prompt particles (or in specialised cases, weak decays of SM particles such as B-hadrons 
or $\tau$ leptons).
Thus in the present analysis, the RH neutrinos are artificially set to be stable, and will manifest themselves as missing transverse energy. 
They will therefore show as missing energy contributions in the fiducial cross sections. 
For low \MNR, this is a good approximation, as seen in Fig.~\ref{fig:neutrinos}a.
For higher neutrino masses, the neutrino should decay, but for most of the parameter space 
the production cross section is very low, as can been seen in Fig.~\ref{fig:neutrinos}b. The exception is at high \gp, but in this region
the contribution from other signatures is also large, and over all the parameter region the contribution to the \textsc{Contur} sensitivity from signatures 
involving heavy neutrinos is negligible.
Taking advantage of displaced vertex signatures is likely to require dedicated searches, and may give additional sensitivity where the 
more conventional signatures fail, as discussed in~\cite{Deppisch:2018eth}.

\subsection{Constraints in \MZP and \gp}

For Case A, the BSM Higgs sector is effectively decoupled by setting $\alpha = 0$, and \MHH is set equal to $\frac{\MZP}{2\gp}$ 
which also ensures vacuum stability (see Section~\ref{sec:vac}). We study the parameter space in \MZP and \gp, with  \MZP and \gp scanned over the ranges
1 GeV $< \MZP < 10$~TeV and $3 \times 10^{-5} < \gp < 0.6$.

\begin{figure}[H]
\vspace{-0.4cm}
\subfloat[]{\includegraphics[width=0.9\textwidth]{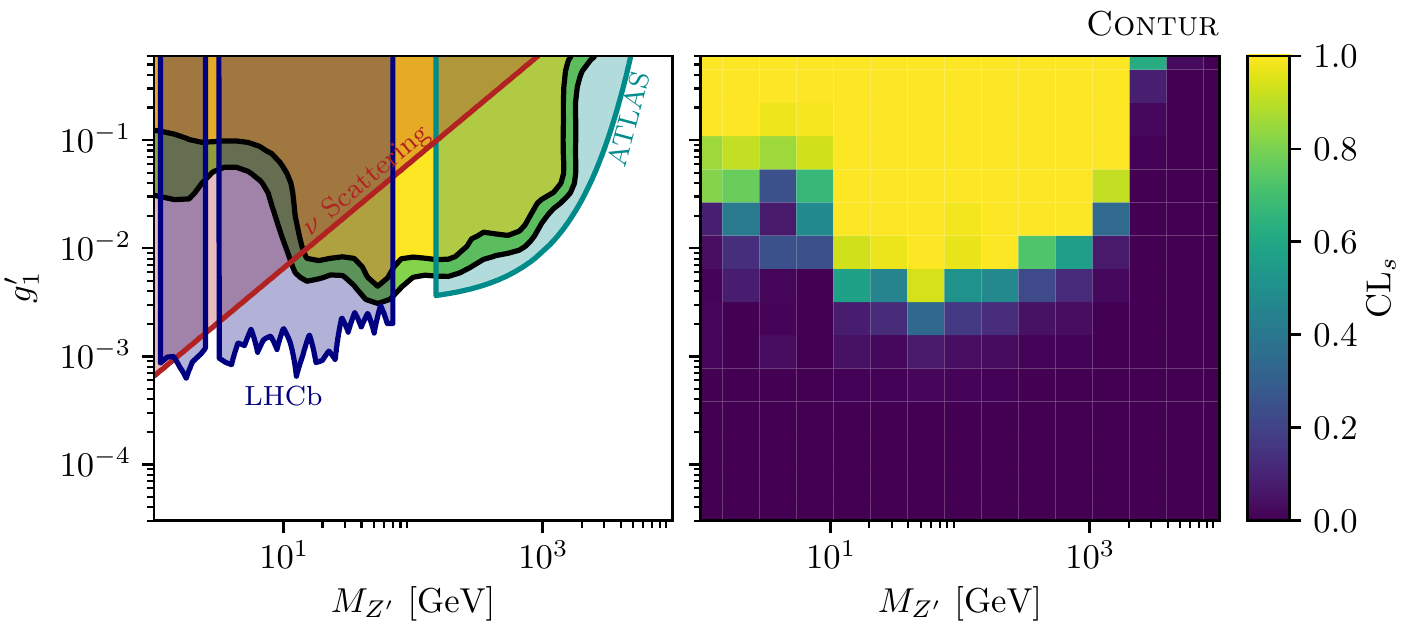}} \vspace{-0.4cm} \\
\subfloat[]{\includegraphics[width=0.9\textwidth]{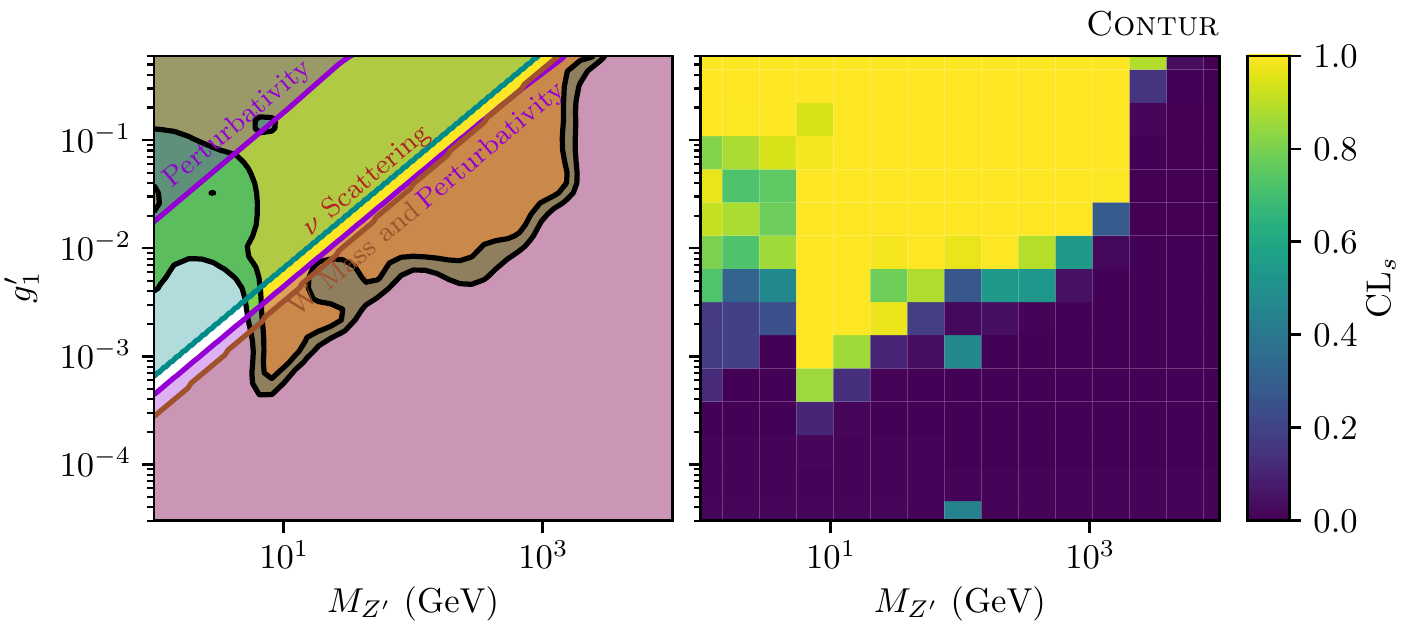}} \vspace{-0.4cm} \\
\subfloat[]{\includegraphics[width=0.9\textwidth]{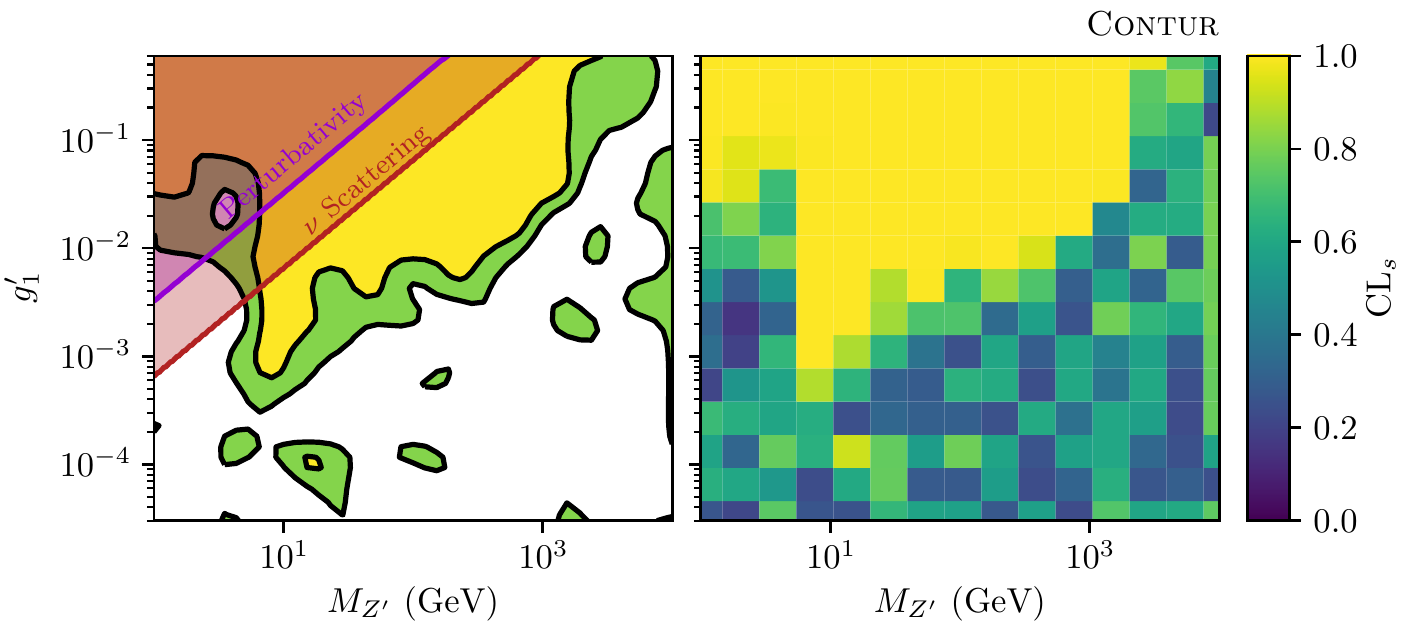}} \vspace{-0.1cm}

\caption{Sensitivity of LHC measurements to the BSM contribution from a gauged B-L model in the \MZP vs \gp plane.
(a) Case A, $\SA = 0$, $\MHH = \frac{\MZP}{2\gp}$;
Left, 95\% (yellow) and 68\% (green) excluded contours. Right, underlying heatmap of exclusion at each scanned parameter space point. 
The 95\% CL limits from the ATLAS search using lepton pairs~\cite{Aaboud:2017buh}, from 
electron-neutrino scattering, from the Darkcast reinterpretation~\cite{Ilten:2018crw} of the LHCb dark photon search~\cite{Aaij:2017rft} and the vacuum stability and perturbativity constraints up to a scale of at least 10~TeV are also indicated;
(b) Case B, $\SA = 0.2, \MHH = \frac{\MZP}{2\gp}$;
as in (a) but for Case B, with additional theory bounds
and constraints from $M_{W}$ and electron-neutrino scattering shown; 
(c) Case C, $\SA = 0.2, \MHH = 200$~GeV;
 as in (b) but for Case C.}
\label{fig:mzgp}
\end{figure}

These settings make our model phenomenologically very similar to the scenario discussed by Batell, Pospelov and Shuve~\cite{Batell:2016zod}, 
and our limits, shown in Fig.~\ref{fig:mzgp}a can be compared to their Fig.~3, as well as to the more recent Fig.~5 of \cite{Ilten:2018crw}. 
In this scenario, the whole plane is allowed by the theoretical constraints 
of Section~\ref{sec:vac}, and the $W$ mass constraint has no impact. 
The LHC data considered in \textsc{Contur} disfavour most of the region for $\gp > 0.01$ for $\MZP < 2$~TeV, and have little sensitivity below this. 
The exclusion comes dominantly from the 
leptonic decays of the \ZP, which would have appeared in various leptonic differential cross sections, and are absent in the data. 
The ATLAS 7 and 8~TeV Drell-Yan measurements~\cite{Aad:2014qja,Aad:2015auj,Aad:2016zzw} have a big impact 
for 12~GeV $< \MZP <$ 1500~GeV, with the $WWW$ cross section~\cite{Aaboud:2016ftt} also having an impact at the highest \MZP.
As expected, our sensitivity tracks that of the ATLAS 13~TeV search, also shown, which  naturally does even better at high \MZP, 
given the higher beam energy. 
No particle-level measurement for this final state in 13~TeV collisions is available in Rivet at time of writing.
The $Z$+jets measurements~\cite{Khachatryan:2014zya,Aaij:2012mda} also disfavour the model in some of this region.
As can been seen, the sensitive region was largely already excluded by the combination of electron-neutrino cross section measurements 
and LHC search data. However, there are regions around $70 < \MZP < 150$~GeV, 
where the limits from current data are weaker (except for a very narrow exclusion around the  $Z$ mass from LEP, not shown).
LHC measurements are able to fill in this window.

Next we consider Case B, in which the \HH mixes with the SM Higgs via a mixing angle of $\sin\alpha = 0.2$. 
The sensitivity plot is shown in Fig.~\ref{fig:mzgp}b.
The electron-neutrino scattering limit still applies, although introduction of an \HH mixing can in principle alter the production
and decay of the \ZP, so the ATLAS dilepton limit does not obviously apply without modification. It can been seen,
however, that the \textsc{Contur} limit derived from the 8~TeV dilepton measurement does not change at high \MZP and high \gp, so it 
is reasonable to assume the ATLAS limit of Fig.~\ref{fig:mzgp}a would also be unchanged.

More significantly, in Case B, the theory constraints come into play, and everything outside the purple coloured lines (i.e. the majority of the parameter plane) 
is ruled out by requiring that the models remain perturbative at least to a scale of 10 TeV, details on the derivation of this are discussed in Section~\ref{sec:benchmarks} and shown explictly for Case B in Fig.~\ref{fig:rgescale}a. This scale choice is displayed as it is deemed to define at least a safe region for the energy scales probed at the LHC. The electron-neutrino scattering limit still applies, and the $W$ Mass constraint also excludes most of the plane for high \MZP
and lower \gp. Only a narrow region remains, along a band from $\MZP \approx 200$~GeV at $\gp = 0.6$ to $\gp \approx 0.002$ at $\MZP =1$~GeV. 
The \textsc{Contur} analysis of LHC data disfavours this entire band.

Cases with fixed \MHH were also considered. If \MHH is set to 1~TeV, the whole plane is excluded if perturbative constraints are applied. 
The experimental sensitivity is very similar to the case where $\MHH = \frac{\MZP}{2\gp}$. For Case C, with $\MHH = 200$~GeV, the theoretically allowed region in \gp and \MZP expands again, such that the only theoretically 
disfavoured region is that already disfavoured by electron-neutrino scattering measurements.
The LHC data disfavour a similar region to Case B, as shown in Fig.~\ref{fig:mzgp}c.
Notable in the heatmap of Fig.~\ref{fig:mzgp}c is the fact there is some sensitivity, albeit weak, over the whole plane.
This is primarily due to the cross section 
for \HH production (via gluon fusion) followed by $h_2$ decay to $WW$, and is thus largely insensitive to $\MZP$ and $\gp$. 
These events make a significant contribution in the phase space of the $l\nu$-jet-jet measurement of \cite{Aaboud:2016ftt}. 
Other signatures involving leptonic decays of $W$ and/or $Z$ also contribute in various regions.
The ATLAS 7~TeV four-lepton measurement~\cite{Aad:2012awa} is particularly important for disfavouring the region where $\MZP$ is 
small and $\gp$ is above $10^{-3}$, since in this region the dominant decay of the $h_2$ is to $\ZP$ pairs and the branching 
fractions $\ZP \rightarrow \mu^+ \mu^-$ and $\ZP \rightarrow e^+ e^-$ are both around 20\%
\footnote{Some of the sensitivity initially seen in this measurement was due to a coding 
bug in the lepton isolation requirement in the relevant \rivet analysis, 
subsequently fixed in \rivet version~3\cite{Bierlich:2019rhm}. This fixed version was used to make these updated plots}.
Although the $\ZP$ is well below the mass window of the $Z$ or even $Z^*$ in the measurement, combinatorials still populate the fiducial region.
Given all of this, one can expect this region to be addressed by future LHC measurements, with increased luminosity and beam energy.

\begin{figure}[h]
\subfloat[]{\includegraphics[width=0.3\textwidth]{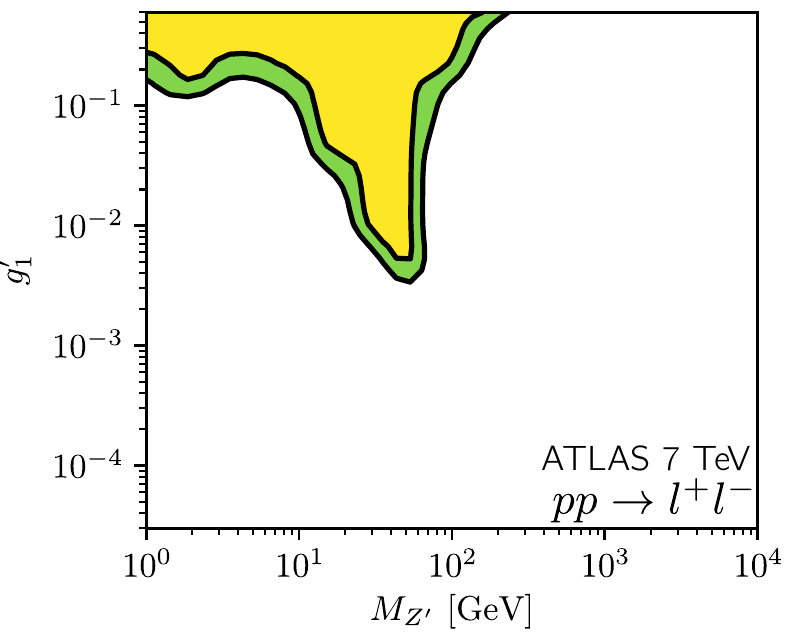}}
\subfloat[]{\includegraphics[width=0.3\textwidth]{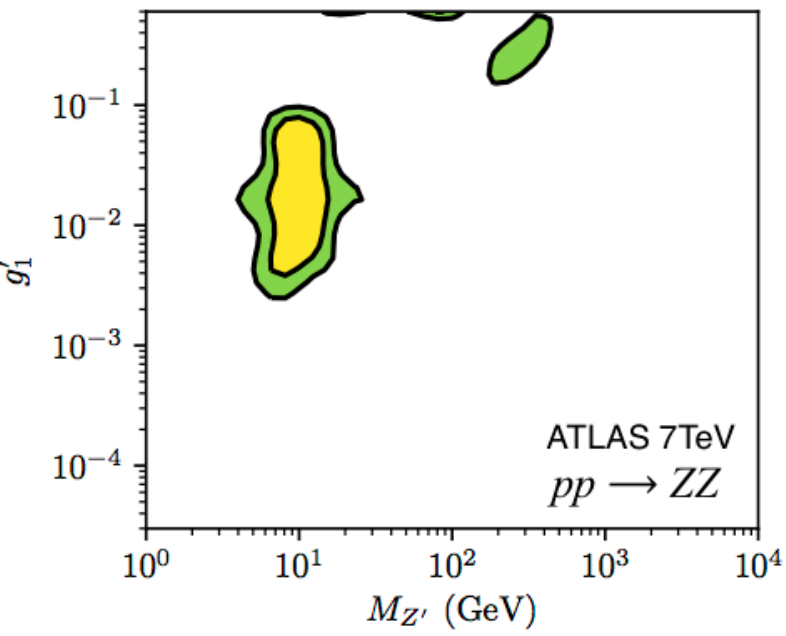}}
\subfloat[]{\includegraphics[width=0.3\textwidth]{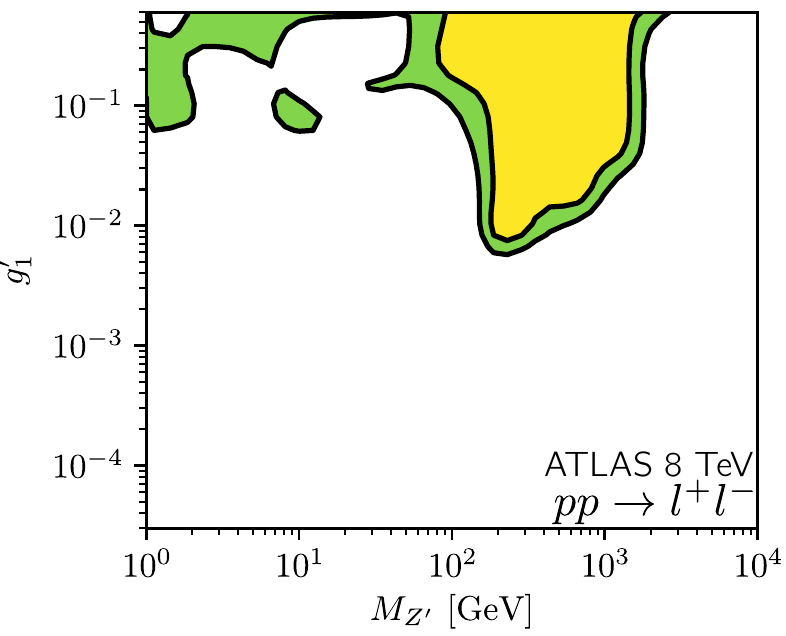}}

\subfloat[]{\includegraphics[width=0.3\textwidth]{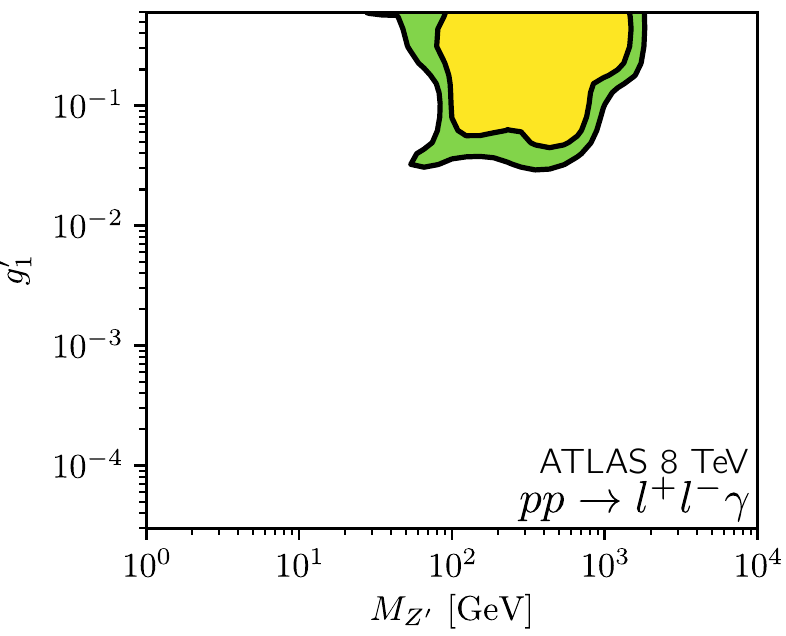}}
\subfloat[]{\includegraphics[width=0.3\textwidth]{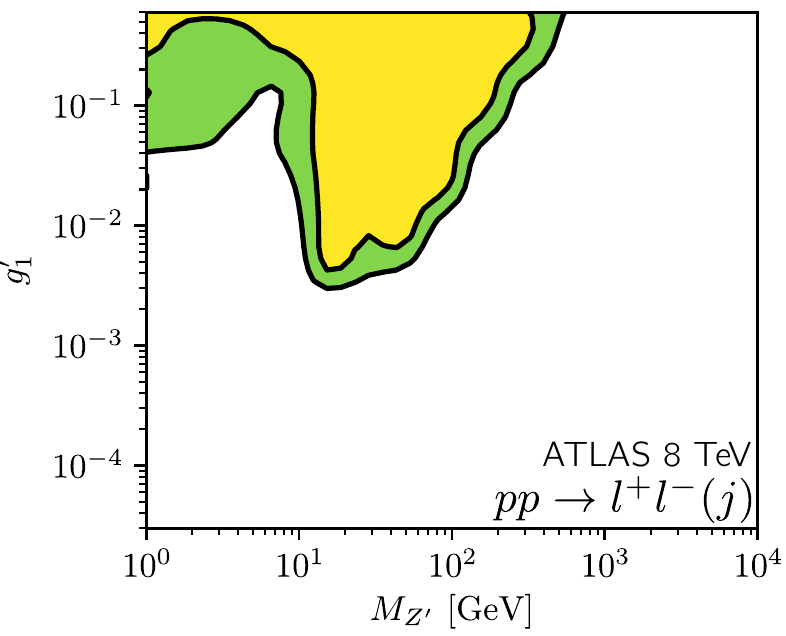}}
\subfloat[]{\includegraphics[width=0.3\textwidth]{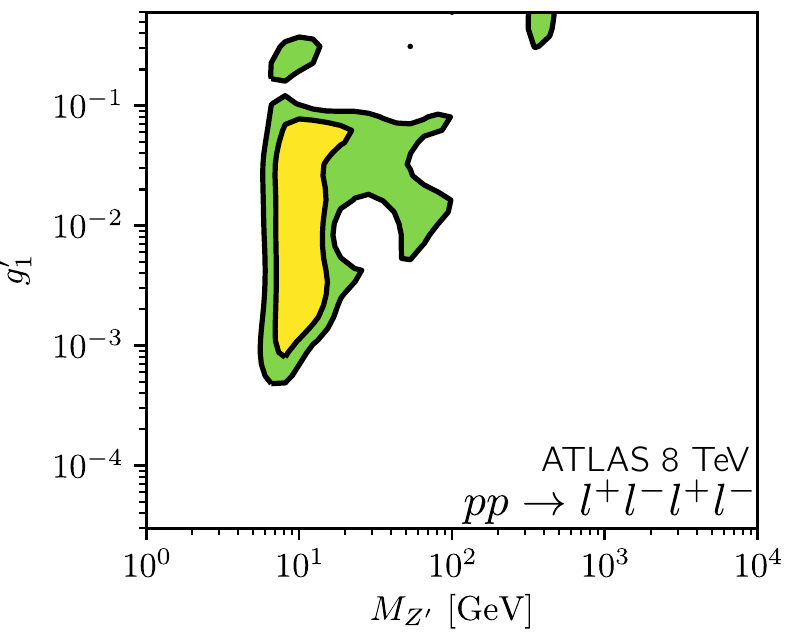}}

\subfloat[]{\includegraphics[width=0.3\textwidth]{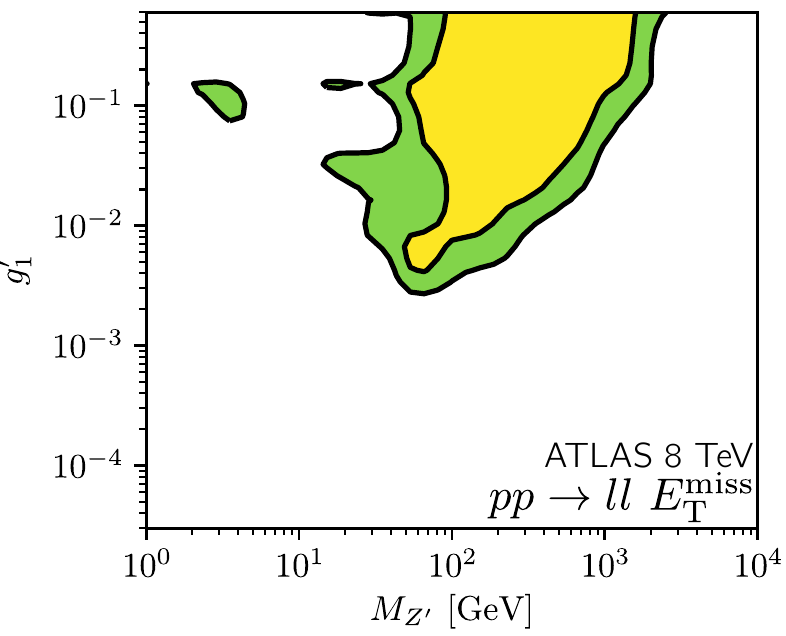}}
\subfloat[]{\includegraphics[width=0.3\textwidth]{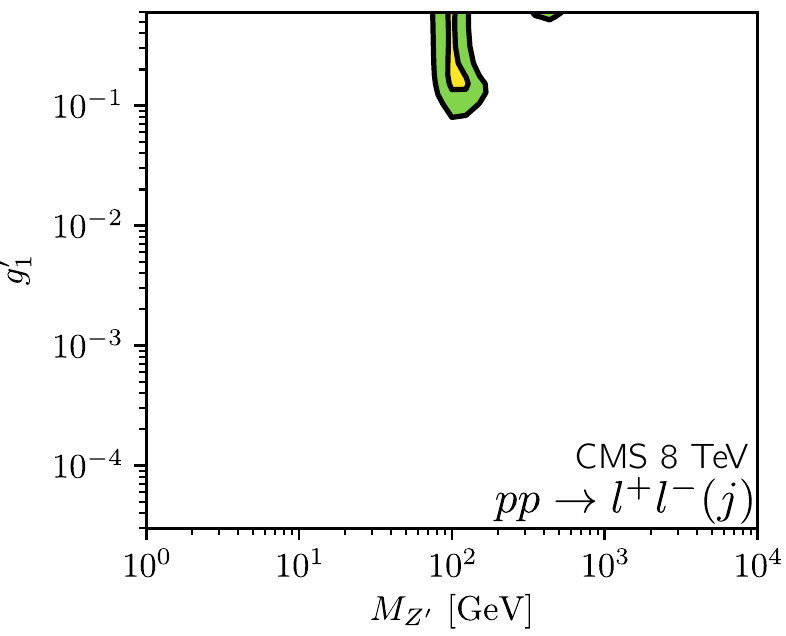}}
\subfloat[]{\includegraphics[width=0.3\textwidth]{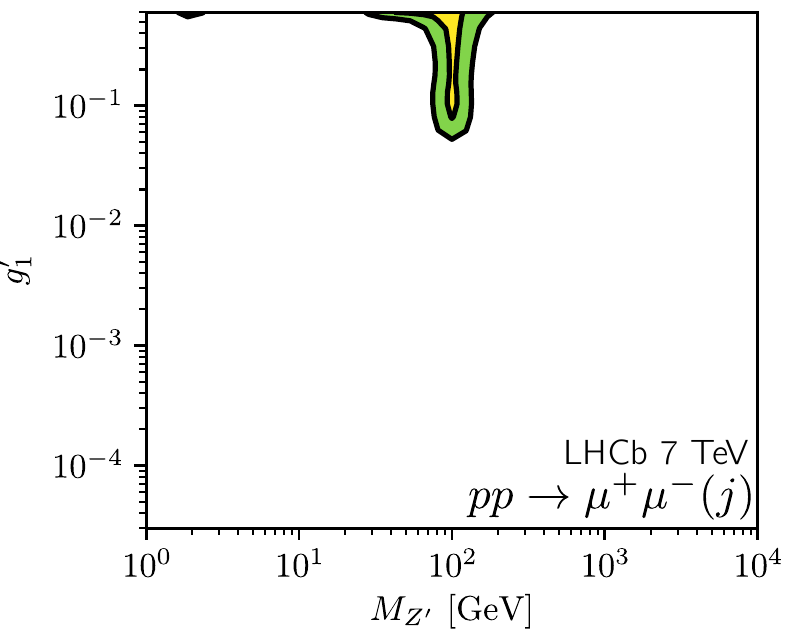}}
\caption{
Disfavoured regions for different, independent measurement classes for Case C.
(a) ATLAS 7 TeV Low mass Drell-Yan measurement~\cite{Aad:2014qja},
(b) ATLAS 7 TeV Four-lepton measurements~\cite{Aad:2012awa},
(c) ATLAS 8 TeV High mass Drell-Yan measurement~\cite{Aad:2016zzw},
(d) ATLAS 8 TeV dilepton plus photon measurements~\cite{Aad:2016sau},
(e) ATLAS 8 TeV Dilepton plus jet measurements~\cite{Aad:2014dta,Aad:2015auj,Aaboud:2017hox},
(f) ATLAS 8 TeV Four-lepton measurements~\cite{Aad:2015rka,Aad:2014tca},
(g) ATLAS 8 TeV Dilepton plus missing transverse energy measurements~\cite{Aaboud:2016ftt,Aad:2016wpd},
(h) CMS 8 TeV dilepton plus jet measurements~\cite{Khachatryan:2016iob},
(i) LHCb 7 TeV dimuon plus jet measurement~\cite{Aaij:2013nxa}.
\label{fig:caseb}}
\end{figure}

As an illustration of how multiple measurements come into play in different regions, in 
Fig.~\ref{fig:caseb} the exclusion bounds for some of the different, statistically independent,
classes of data used by \textsc{Contur} are shown.
It may be seen that the measurements sensitive to the \ZP still play a role, but the sensitivity is extended to lower values,
and the four-lepton~\cite{Aaboud:2017rwm,Aad:2014tca,Aad:2012awa} measurements contribute to this (see e.g. Fig.~\ref{fig:caseb}b) 
The main change in this scenario is that for low \MZP, the decay branching ratio of the both the SM Higgs and \HH 
to $\ZP$ is significant, and the leptons from the $\ZP$ decay appear in the fiducial phase space of several measurements. 
Of course, the ATLAS and CMS measurements of Higgs properties,
which are not used by \textsc{Contur}, would also rule out some of these scenarios.

\begin{figure}[h]
\begin{tabular}{cc}
\subfloat[]{\includegraphics[width=0.45\textwidth]{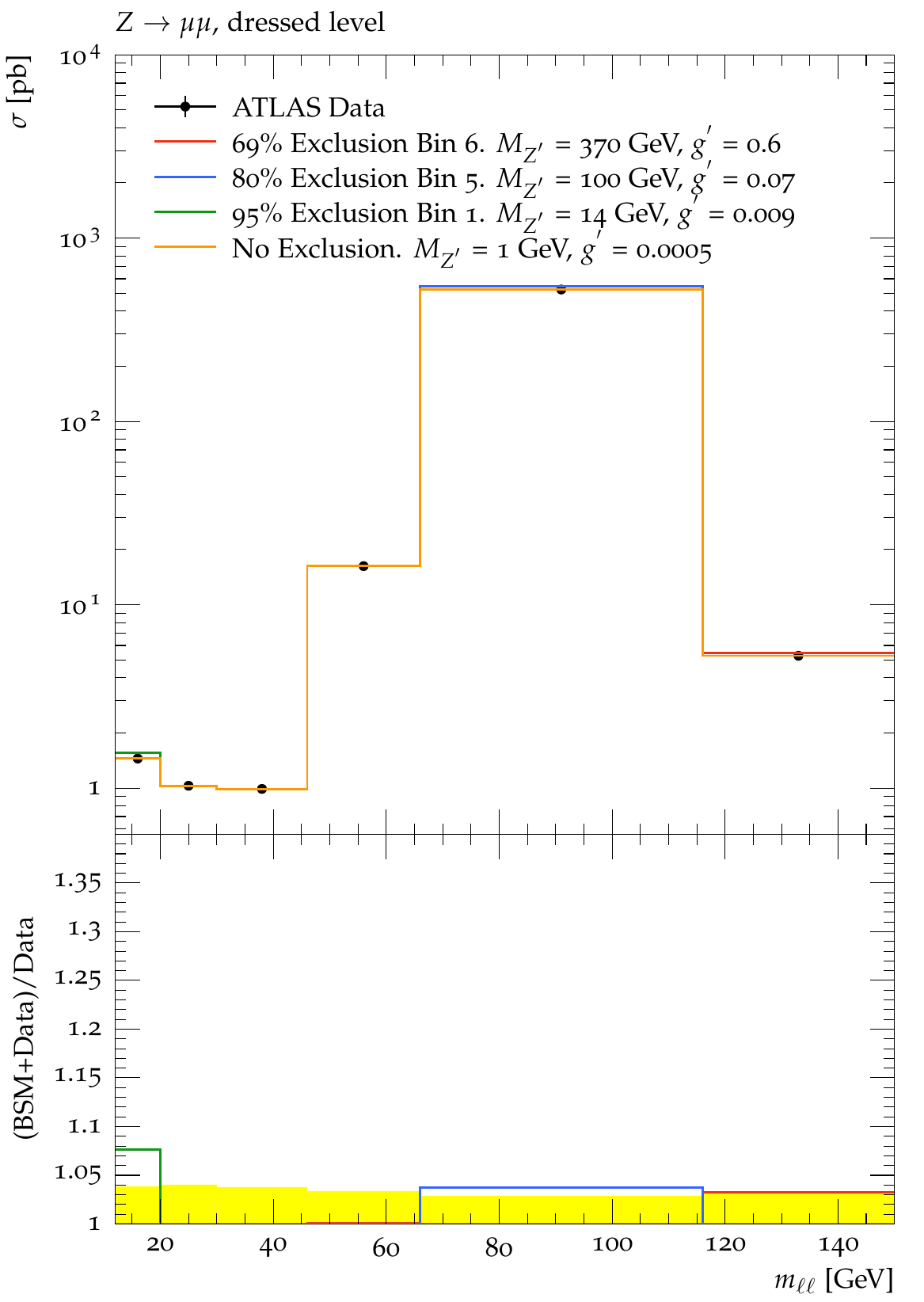}}
&
\shortstack{
\subfloat[]{\includegraphics[width=0.45\textwidth]{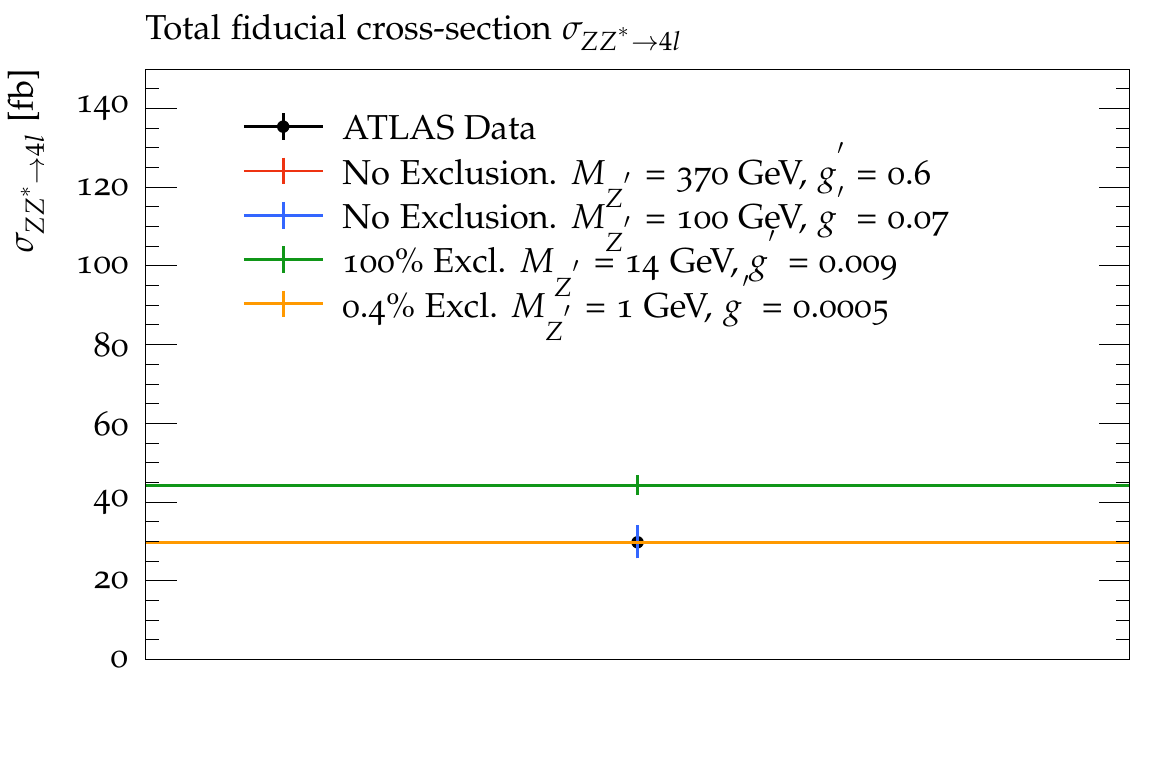}} \\
\subfloat[]{\includegraphics[width=0.45\textwidth]{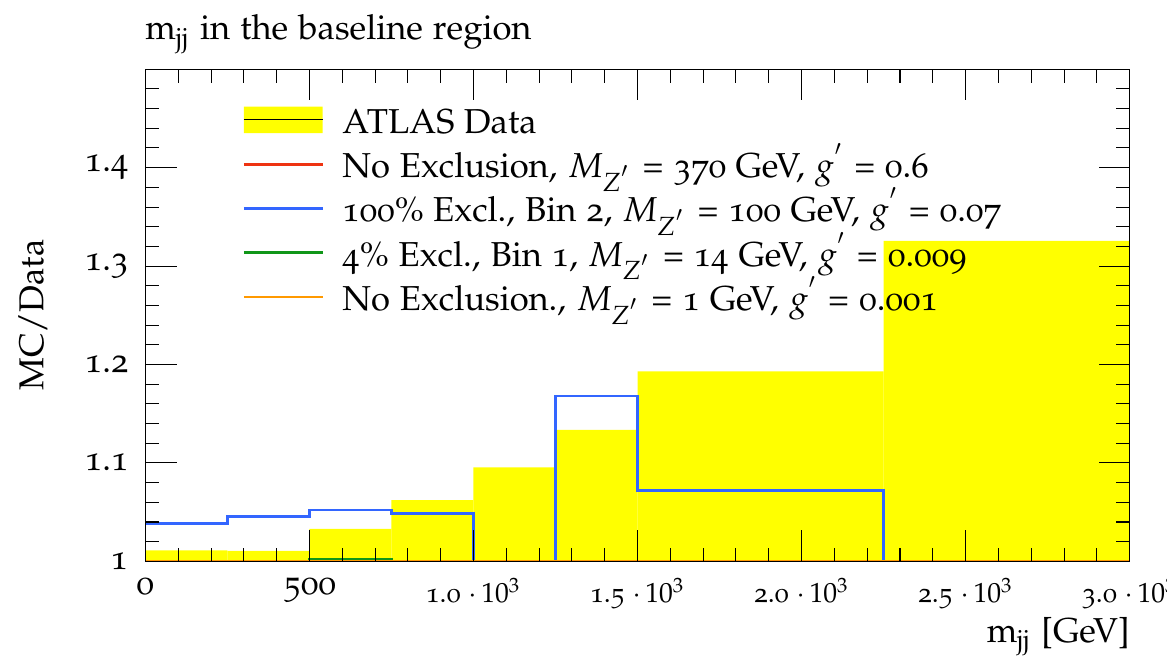}}}
\end{tabular}
\caption{Examples of the exclusion from four points in the parameter space moving along the below
the region of Case C excluded by neutrino scattering, Fig.\ref{fig:mzgp}c.
(a) The dimuon mass measurement from \cite{Aad:2015auj},
(b) The $ZZ^*$ (four lepton) measurement from \cite{Aad:2012awa}, 
(c) The dijet mass in $Z$ events from \cite{Aad:2014dta}.
The legend indicates the parameter point in \MZP and \gp space and the bin of the 
plot which gives the sensitivity. 
$\MHH = 200$ GeV
}
\label{fig:riveteg1}
\end{figure}

Going even further into detail, Fig.~\ref{fig:riveteg1} shows examples of some of the distributions which have exclusion 
power for this scenario. In \ref{fig:riveteg1}a, single \ZP production dominates, except for the highest 
mass point where combinatorials from multilepton events from \ZP and \HH production contribute. In \ref{fig:riveteg1}b, for low \MZP, pair production (including from \HH decays) dominates, while for higher 
\MZP, \ZP + \HH production contributes. In \ref{fig:riveteg1}c, there is a powerful exclusion when 
\MZP is within the $Z$ mass window of the analysis.
Other data, including the dimuon mass, but also for example the 8~TeV $Z+\gamma$ results~\cite{Aad:2016sau}, 
see Fig.~\ref{fig:caseb}d, 
also play a role.
The cross sections and branching ratios calculated by Herwig for the most important processes are given in
Table~\ref{tab:processes1} for each of the parameter points of Fig.~\ref{fig:riveteg1}.

\begin{table}[h]
\centering	
\begin{tabular}{|c|c||c|c|c|c|c|}
\hline
\MZP  & \gp    & Production              & Cross Section  & Decay                    & Branching    \\  
(GeV) &        & Process                 & ($\sigma$, pb) &                          & Fraction     \\  
\hline 
\hline 
1     & 0.0005 & $gg \rightarrow \ZP\ZP$ & 0.6            & $\ZP \rightarrow l^+l^-$ & 0.36          \\
      &        & $gg \rightarrow g \HH$  &  0.078            & $\HH \rightarrow \ZP\ZP$ & 0.58          \\
\hline
14    & 0.009  & $u\bar{u} \rightarrow g\ZP$ & 40.6    & $\ZP \rightarrow l^+l^-$ & 0.27         \\
\hline
100   & 0.07   & $u\bar{u} \rightarrow \ZP \rightarrow l^+l^-$ & 31 & $\ZP \rightarrow l^+l^-$ & 0.27 \\
\hline
370   & 0.6    & $u\bar{u} \rightarrow \ZP \rightarrow l^+l^-$ & 30 & $\ZP \rightarrow l^+l^-$ & 0.27  \\
\hline
\end{tabular}
\caption{Cross sections (in 8~TeV $pp$ collisions) and branching fractions for the main processes contributing to Fig.~\ref{fig:riveteg1}.}
\label{tab:processes1}
\end{table}

\subsection{Constraints in \MHH and $\sin{\alpha}$}

\begin{figure}[h]
  \subfloat[]{\includegraphics[width=1.0\textwidth]{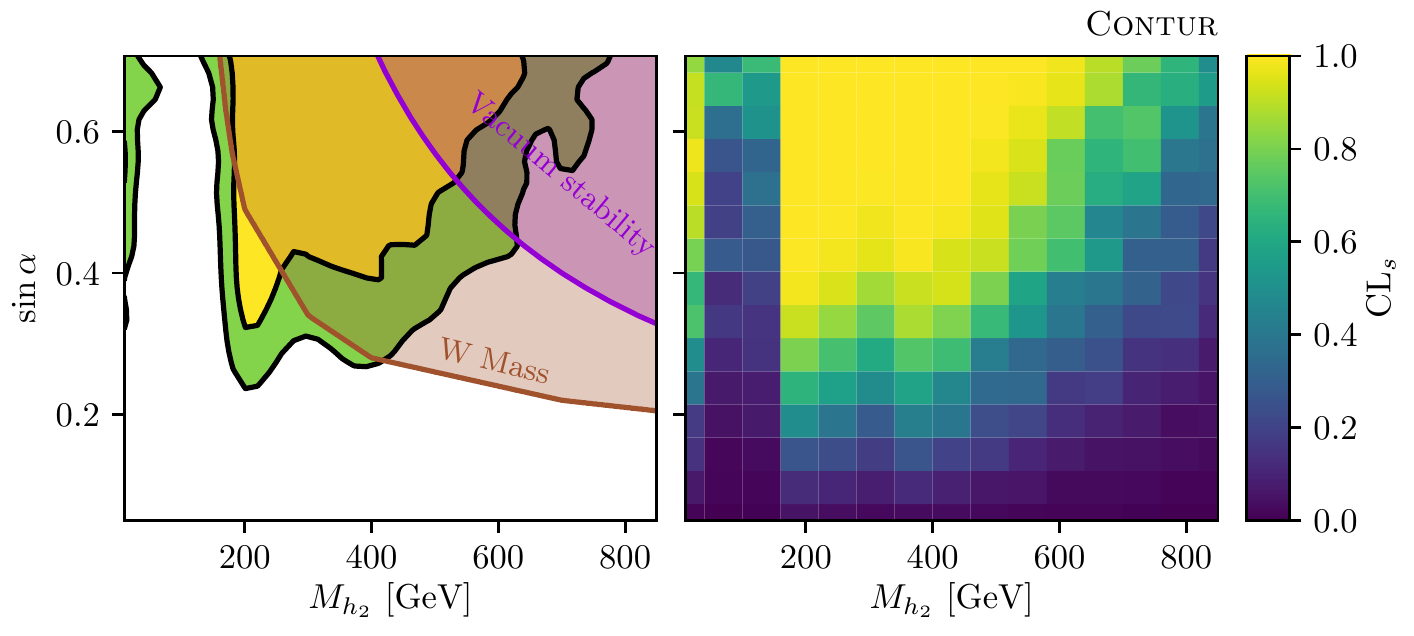}} \\
  \subfloat[]{\includegraphics[width=1.0\textwidth]{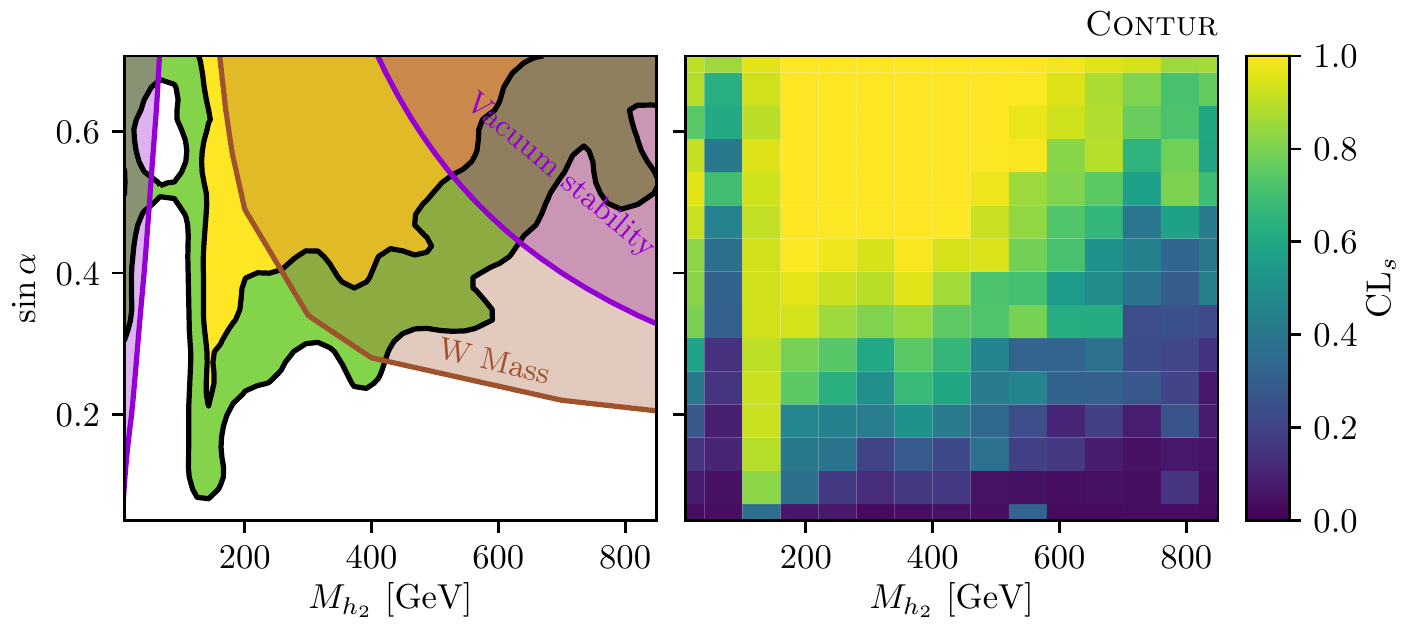}}
  
\caption{Sensitivity of LHC measurements to the BSM contribution from a gauged B-L model in the \MHH vs \SA plane,
(a) Case D, \gp = 0.2, \MZP = 7~TeV.
  Left, 95\% (yellow) and 68\% (green) excluded contours. Right, underlying heatmap of exclusion at each scanned parameter space point. The theory constraints from perturbativity and vacuum stability, requiring the model to be well behaved up to at least 10~TeV, as well as the constraint from $M_W$ are also shown.
(b) Case E, \gp = 0.001, \MZP = 35~GeV. Figures as in (a) but for Case E.
}
\label{fig:mh2sa1}
\end{figure}

Another interesting possibility is Case D, where the \ZP is heavy, and thus decouples, but the \HH mixes significantly with the SM Higgs. 
In this case the mixing has a negligible effect on the SM Higgs branching ratios, and the sensitivity of the cross section measurements used by
\textsc{Contur} relies upon \HH production.
Figure~\ref{fig:mh2sa1}a shows the plane in \MHH and \SA. 
The upper right portion of the plane is excluded by constraints on $M_W$, while perturbativity constraints, requiring the model to be perturbative and stable up to at least 10~TeV, eliminate a smaller region in the top right.
The LHC measurements have some sensitivity at larger mixing angles ($\SA \ge 0.4$), 
centred on the $\HH \rightarrow WW, ZZ$ threshold at $\approx 200$~GeV, and the
$\HH \rightarrow t\bar{t}$ threshold at around 400~GeV. 
The most sensitive measurements here are the two-lepton-plus-two-jet cross section from~\cite{Aaboud:2016ftt}
and the four-lepton cross section of~\cite{Aaboud:2017rwm}.
The heat map in Fig.~\ref{fig:mh2sa1}a indicates that LHC data do have some sensitivity reach at other \MHH values 
and lower \SA, so that more of the parameter space is likely to become accessible if the precise measurements are made. 
Such measurements should be made as the LHC accumulates more integrated luminosity.

In some sense, Case E is complementary, and is shown in Fig.~\ref{fig:mh2sa1}b. 
\MZP is now low, but \ZP production is now suppressed by
fixing a low value of \gp. The exclusion derived from \textsc{Contur} is similar to Case D, but varies at lower
\MHH values as the decay $\HH \rightarrow \ZP \ZP$ can continue to have an impact to lower \MHH. 

\begin{figure}[h]
\subfloat[]{\includegraphics[width=0.45\textwidth]{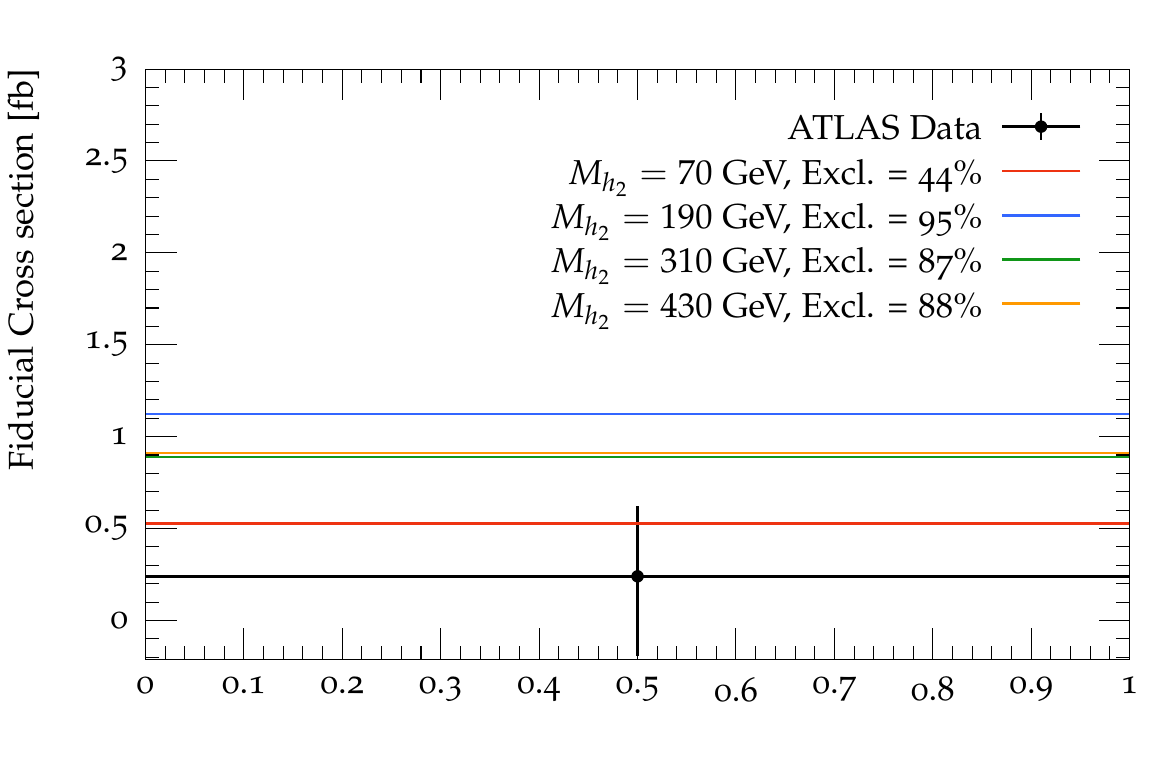}\label{fig:2l2j}}
\subfloat[]{\includegraphics[width=0.45\textwidth]{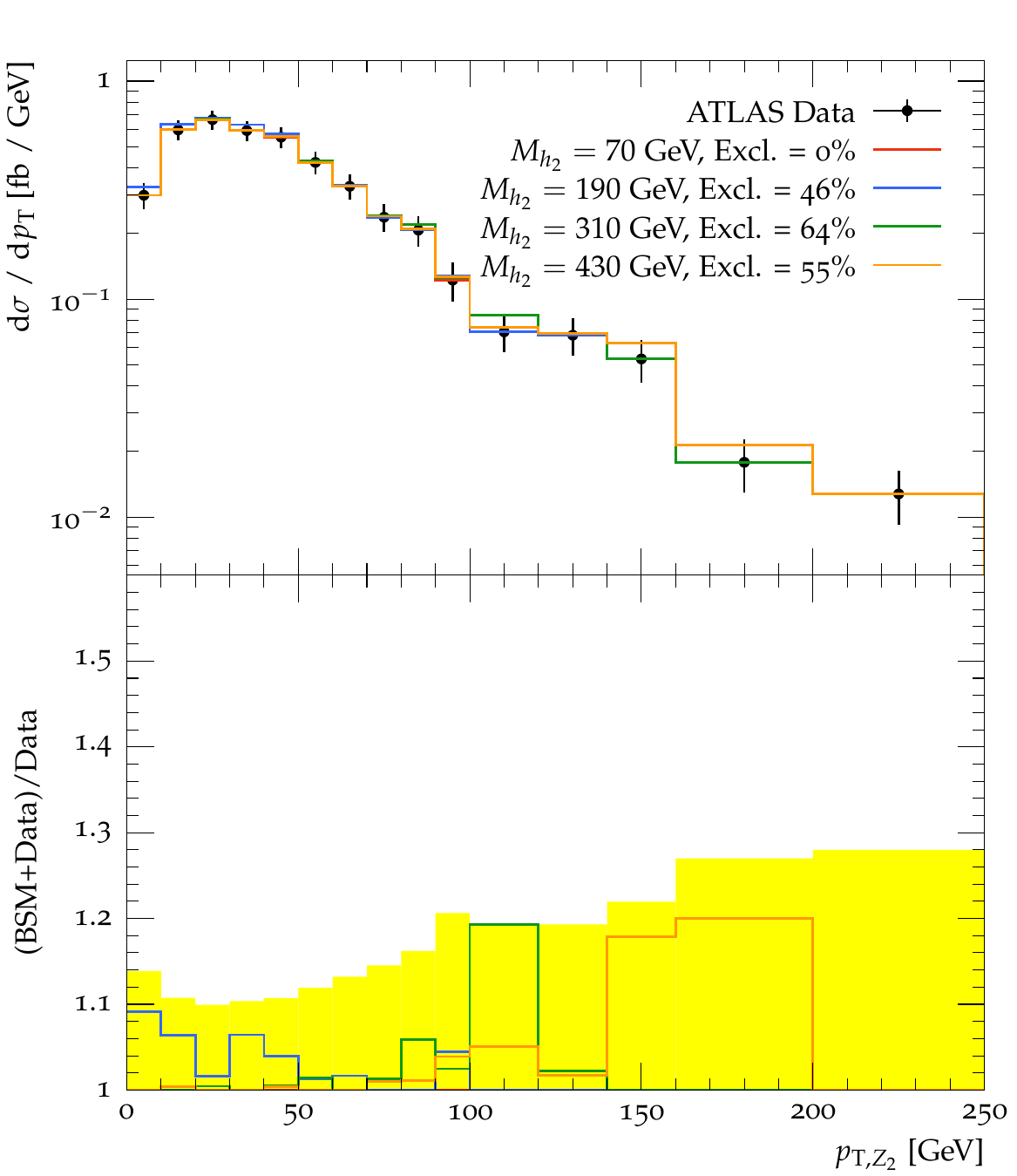}\label{fig:ZZ13}}
\caption{Examples of the exclusion from four points in the parameter space moving along the lower 
edge of the theoretically 
allowed region of Fig.\ref{fig:mh2sa1}a.
(a) The dilepton plus dijet measurement from \cite{Aaboud:2016ftt},
(b) The $ZZ^*$ (four lepton) measurement from \cite{Aaboud:2017rwm}, 
The legend indicates the parameter point in 
\MZP = 7 TeV, \gp = 0.2, \SA = 0.42
}
\label{fig:riveteg2}
\end{figure}

The most important processes contributing for these parameter points are summarised in Table~\ref{tab:processes2}. 
Finally it should be noted that when \MHH is near or below the SM Higgs mass, there will be a significant impact 
on SM Higgs signatures which are not considered in the \textsc{Contur} analysis but which are likely to disfavour much of
the parameter space for $\MHH \lessapprox 150$~GeV when the Higgs mixing is significant. 

\begin{table}[h]
\centering	
\begin{tabular}{|c||c|c|c|c|c|c|}
\hline
\MHH  &  Production              & Cross Section  & Decay                      & Branching    \\  
(GeV) &  Process                 & ($\sigma$, pb) &                            & Fraction     \\  
\hline 
\hline 
70    & $u\bar{u} \rightarrow Z\HH$ & 0.13        & $\HH \rightarrow b\bar{b}$ & 0.88          \\
\hline
190   & $gg \rightarrow g\HH$       & 0.37        & $\HH \rightarrow WW$       & 0.78          \\
      &                             &             & $\HH \rightarrow ZZ$       & 0.21          \\
\hline
310   & $gg \rightarrow g\HH$       & 0.20        & $\HH \rightarrow WW$       & 0.51         \\
      &                             &             & $\HH \rightarrow ZZ$       & 0.27          \\
      &                             &             & $\HH \rightarrow hh$       & 0.22          \\
\hline
430   & $gg \rightarrow g\HH$       & 0.14        & $\HH \rightarrow WW$       & 0.46         \\
      &                             &             & $\HH \rightarrow ZZ$       & 0.22          \\
      &                             &             & $\HH \rightarrow hh$       & 0.21          \\
      &                             &             & $\HH \rightarrow t\bar{t}$ & 0.11          \\
\hline
\end{tabular}
\caption{Cross sections (in 8~TeV $pp$ collisions) and branching fractions for the main processes contributing to Fig.~\ref{fig:riveteg2}.}
\label{tab:processes2}
\end{table}

\section{Conclusions}
\label{sec:conclusions}

The new particles and interactions implied by a model based on gauging the baryon number minus lepton number $B-L$ symmetry
are simulated across a wide range of parameter space for proton-proton collisions at the LHC. For significant regions of parameter
space, the new interactions contribute to signatures and phase space in which LHC measurements have already been made and in which
the data have been shown to agree well with the SM. Thus in these regions the model is disfavoured or excluded already.

When the exotic Higgs of the model (\HH) is decoupled, the phenomenology is rather simple and the main sensitivity comes from the 
production of the new gauge boson, \ZP and its decays to leptons. In this case,
our results at high \MZP reproduce those of resonance searches made by ATLAS and CMS in the same data set. 
At lower \MZP and for couplings \gp greater than about $7 \times 10^{-3}$, some previously unexamined parameter space is excluded compared to
the summary of~\cite{Batell:2016zod}, in the region around $10 < \MZP < 30$~GeV and around the $Z$ mass.

If the exotic Higgs sector mixes with the SM Higgs, with a mixing angle $\alpha$, we show that the sensitivity in the LHC data at high \MZP is retained,
and that the sensitivity now extends to lower \gp; for $\SA \gtrapprox 0.2$ the model is disfavoured for values of \gp above about $5 \times 10^{-3}$ 
for a wide range of \MHH. This extension is driven by the decays of the \HH, principally to $W$ bosons, 
although for low \MZP the decays of the SM Higgs to \ZP pairs are also important, and considering both these channels and the \ZP decays to leptons
in combination gives a more powerful limit than previously obtained. 
This is only possible because of the wide array of experimental signatures which can be considered in parallel using \textsc{Contur}.

If the \ZP is suppressed either because \MZP is high or \gp is low, the sensitivity comes entirely from the extended Higgs sector.
The limit on \SA for this specific model using existing measurements is similar to that obtained by combining Higgs searches and Higgs signal rates in 
general extended scalar-sector models~\cite{Dev:2012bd}. 
For $\SA < 0.2$ and $\gp < 5 \times 10^{-3}$, substantial regions of parameter space remain open, even for low \MZP.

Some sensitivity, below 95\% exclusion, is seen at lower \SA and higher \MHH values. 
The studies presented here use only the relatively small fraction of LHC data currently available as fiducial, 
particle-level measurements in HEPDATA and Rivet. 
As more data are collected, and increasingly precise measurements are made available in this manner, 
the sensitivity will grow into these further regions.

\acknowledgments
This work has received funding from the European Union's Horizon 2020 research and innovation programme as part of the 
Marie Sklodowska-Curie Innovative Training Network MCnetITN3 (grant agreement no. 722104).
We thank Ben Waugh for useful discussions early on in the project, 
and Peter Richardson for improvements and enhancements of the Herwig UFO interface.

\input appendix.tex
\bibliographystyle{JHEP}
\bibliography{B-L-contur}

\end{document}

%% file: appendix.tex
\newpage

\section{Appendices}

\subsection{Renormalization Group Equations}
\label{sec:app}
The Renormalisation Group Equations (RGEs) for the minimal $SU(3) \times SU(2)_L \times U(1)_Y \times U(1)_{B-L}$ model are given in \cite{Chakrabortty:2013zja}. We here list the relevant RGEs for the convenience of the reader. The RGEs for the gauge coupling constants $g_1$ (associated with $U(1)_Y$), $g$ ($SU(2)_L$), $g_s$ ($SU(3)$) and $g_1^\prime$ ($U(1)_{B-L}$) are given by
\begin{align}
\label{RGEgauge}
	16 \pi^2 \frac{d}{dt} g_1 &= \frac{41}{6}{g_1}^3,  \\
	16 \pi^2 \frac{d}{dt} g   &= -\frac{19}{6}{g}^3, \\
	16 \pi^2 \frac{d}{dt} g_s &= -7 g_s^3, \\
	16 \pi^2 \frac{d}{dt} g_1^{\prime}  &= 12 {g_1^{\prime}}^3 
	+ \frac{32}{3} g_1^{\prime} \tilde{g}+\frac{41}{6} g_1^{\prime} {\tilde{g}}^2,  \\
	16 \pi^2 \frac{d}{dt} \tilde{g}  &= \frac{41}{6} \tilde{g}({\tilde{g}}^2+2g_1^2)+\frac{32}{3}g_1^{\prime}({\tilde{g}}^2+g_1^2)+12 {g_1^{\prime}}^2 \tilde{g}.
\end{align} 
The last line describes the mixing between the $U(1)$ terms $U(1)_Y$ and $U(1)_{B-L}$. 

In the Yukawa sector, we only include the effect of the large top quark Yukawa coupling $Y_t$ and the (potentially large) Yukawa coupling $y^M_{ij}$ of the right-handed neutrino to the singlet scalar $\chi$. They are given by
\begin{align}
\label{RGEyukawa}
	16 \pi^2 \frac{d}{dt} Y_t  &= Y_t[\frac{9}{2}Y_t^2 
	- 8 g_s^2 - \frac{9}{4}g^2 - \frac{17}{12}g_1^2 
	- \frac{17}{12}{\tilde{g}}^2 - \frac{2}{3}g_1^{\prime} 
	- \frac{5}{3}\tilde{g}g_1^{\prime}] \\
	16 \pi^2 \frac{d}{dt}y_i^M  &= y_i^M [4(y_i^M)^2 + 2Tr[(y^M)^2]-6g_1^{\prime}],
\end{align}
where, for simplicity, we assume diagonal right-handed neutrino Yukawa couplings $y^M_{ij} = y^M_i\delta_{ij}$.

Finally, the RGEs for the couplings in the scalar sector, $\lambda_1$, $\lambda_2$, $\lambda_3$, are given by
\begin{align}
\label{RGEscalar}
	16 \pi^2 \frac{d}{dt} \lambda_1 & = 
	24\lambda_1^2 + \lambda_3^2 - 6Y_t^4 + \frac{9}{8}g^4 
	+ \frac{3}{8}g_1^4 
	+ \frac{3}{4}g^2 g_1^2 
	+ \frac{3}{4}g^2 \tilde{g}^2 
	+ \frac{3}{4}g_1^2 \tilde{g}^2, \nonumber\\
	& + \frac{3}{8} \tilde{g}^4 + 12 \lambda_1 Y_t^2 - 9\lambda_{1}g^2 - 3\lambda_1 g_1^2 - 3\lambda_1 \tilde{g}^2, \\
	8 \pi^2 \frac{d}{dt} \lambda_2 &= 
	10\lambda_2^2 + \lambda_3^2 - \frac{1}{2}\text{Tr}[(y^M)^4] 
	+ 48{\gp}^4 
	+ 4 \lambda_2 \text{Tr}[(y^M)^2] - 24\lambda_2 {\gp}^2, \\
	8 \pi^2 \frac{d}{dt} \lambda_3 & = 
	\lambda_3 \left(6\lambda_1 + 4\lambda_2 + 2\lambda_3 + 3Y_t^2 
	- \frac{3}{4}(3g - g_1^2 - \tilde{g}^2) + 2\text{Tr}[(y^M)^2] 
	- 12 {\gp}^2\right) + 6\tilde{g}^2{\gp}^2.
\end{align} 